# A BIBLIOMETRIC ANALYSIS OF PUBLISHER AND JOURNAL INSTRUCTIONS TO AUTHORS ON GENERATIVE-AI IN ACADEMIC AND SCIENTIFIC PUBLISHING

Conner Ganjavi[1,2], Michael B. Eppler, [1,2] Asli Pekcan[1,2], Brett Biedermann, [1,2] Andre Abreu[1,2], Gary S. Collins[3], Inderbir S. Gill[1,2], Giovanni E. Cacciamani[1,2,*]

*[1] USC Institute of Urology and Catherine and Joseph Aresty Department of Urology, Keck School of Medicine, University of Southern California, Los Angeles, California.*
*[2]Artificial Intelligence Center at USC Urology, USC Institute of Urology, University of Southern California, Los Angeles, California.*
*[3] UK EQUATOR Centre, Centre for Statistics in Medicine, Nuffield Department of Orthopaedics, Rheumatology and Musculoskeletal Sciences, University of Oxford, Oxford, UK.*

Funding: None
Declaration of conflicts of interest: None
Wordcount: 3791
Figures: 1
Tables: 2
Supplemental materials: 0
References: 26

**Corresponding Author***
*Giovanni E. Cacciamani, MSc, MD, FEBU
Assistant Professor of Research Urology
Assistant Professor of Research Radiology
Director Artificial Intelligence Center at USC Urology
USC Institute of Urology Catherine and Joseph Aresty Department of Urology
Keck School of Medicine, University of Southern California, Los Angeles, CA, USA.
Email: Giovanni.cacciamani@med.usc.edu
Phone: +1 (626) 4911531P

**ABSTRACT**

**Objectives** To determine the extent and content of guidance for authors regarding the use of generative-AI (GAI) among the top 100 academic publishers and journals in science.

**Design** Cross-sectional, bibliometric study

**Setting** The publishers were identified by the total number of journals. The journals were identified via SCImago Journal Rank using the H-index as an indicator of journal productivity and impact.

**Participants** 100 publishers and journals in science were included in this study regardless of subject, language, or country of origin. The websites of these publishers and journals were screened from between 19th and 20th May 2023.

**Main Outcome Measures** Descriptive statistics was used to characterize the prevalence and content of GAI guidance listed on publisher or journal websites, and to analyze the consistency of guidance between publishers and affiliated journals.

**Results** Among the largest 100 publishers, 17% provided guidance on the use of GAI, of which 12 (70.6%) were among the top 25 publishers. Among the top 100 journals, 70% have provided guidance on GAI. Of those with guidance, 94.1% of publishers and 95.7% of journals prohibited the inclusion of GAI as an author. Four journals (5.7%) explicitly prohibit the use of GAI in the generation of a manuscript, while 3 (17.6%) publishers and 15 (21.4%) journals indicated their guidance exclusively applies to the writing process. When disclosing the use of GAI, 42.8% of publishers and 44.3% of journals included specific disclosure criteria. There was variability in guidance of where to disclose the use of GAI, including in the methods, acknowledgments, cover letter, or a new section. There was also variability in how to access GAI guidance and the linking of journal and publisher instructions to authors. Two journals had GAI guidance that directly conflicted with guidance developed by their publishers.

**Conclusions** There is a lack of guidance by some top publishers and journals on the use of GAI by authors. Among those publishers and journals that provide guidance, there is substantial heterogeneity in the allowable uses of GAI and in how it should be disclosed, with this heterogeneity persisting among affiliated publishers and journals in some instances. The lack of standardization burdens authors and threatens to limit the effectiveness of these regulations. There is a need for standardized guidelines in order to protect the integrity of scientific output as GAI continues to grow in popularity.

## INTRODUCTION

In recent years, advances in artificial intelligence (AI) have spurred the creation of many AI-based tools for use in research[1-3]. Generative-AI (GAI) utilizes Large Language Models (LLMs) to generate unique text or image-based responses to user prompts, and has gained popularity since the release of Generative Pre-trained Transformers (GPT), namely ChatGPT by OpenAI on Nov 30th, 2022[4]. Within two months, ChatGPT reached 100 million monthly users, making it the fastest technology adoption in history at the time[5]. Now other products from major technology companies like Google's Bard and MedPalm or Microsoft's Bing Chat[6-8] are developing quickly, with a technological uptake never seen before.

Naturally, the advent of this new technology has resulted in a substantial upsurge of academic interest, accompanied by a pronounced acceleration in its potential utilization. To date, there have been over 650 different research articles and editorials discussing the applications and pitfalls of GAI, many of which use GAI within the research and writing process itself. Regarding use in research and academic writing, studies frequently mention GAI's ability to improve grammar and vocabulary[9], translate text into various languages[10], propose novel research ideas[9], synthesize large amounts of information[11], suggest statistical tests[12], write code and novel textual content[10 12], and streamline the overall research process[13]. However, authors have been warned that GAI cannot be held accountable for its output, which has a risk of inaccuracy, bias, and plagiarism among other pitfalls[11 13]. Given these concerns and GAI's rapid adoption, publishers and journals have responded quickly to develop guidance on proper use.

On December 9th, 2022, *Nature* published the first paper discussing concerns about the use of ChatGPT and GAI in academic writing[14]. Since then, journals and publishers have begun updating their editorial policies and instruction to authors to provide guidance on how to disclose the use of GAI in academic research. The journal *Science* published an article on January 27th, 2023 stating their decision to prohibit the use of GAI to generate text, figures images or graphics in the writing process, and views violation of the policy as constituting scientific misconduct [15].

Other journals have allowed its use with restrictions and a require full disclosure[16]. The Committee on Publication Ethics (COPE) released a position statement on AI tools in research publications on February 13th, 2023[17] emphasizing that *“ [..] AI tools cannot meet the requirements for authorship as they cannot take responsibility for the submitted work”* while also suggesting ways to disclose AI use and emphasizing that authors are accountable for the work produced by AI tools[17]. Even if the current COPE AI statement is promptly endorsed by journals (e.g., *JAMA*[18 19]) and editorial associations (e.g., WAME[20]), it does not provide a comprehensive and functional set of recommendations on key aspects to guide responsible GAI tool usage in scientific writing. Specifically, it fails to address certain potential pitfalls of these tools, does not offer a standard disclosure statement detailing specific elements to be included. This gap in standardization led to a variety of bespoke guidance formulated by individual journals and publishers for addressing AI usage in scientific publications[21].

In this study, our aim is to examine the extent and nature of author guidelines pertaining to GAI usage across the largest 100 academic publishers and scientific journals. Our objective is to identify the shared characteristics, any methodological details on how guidance was developed, as well as the variations in the guidance, with the goal of assessing their commonalities and divergences.

## METHODS

### *Publisher Selection and Data Acquisition*

We utilized the list provided in the study by Nishikawa-Pacher[22] which identified and ranked the largest 100 publishers by journals count. The largest publisher on the list comprises 3763 journals, while the smallest on the list publishes 76 journals. In total, these 100 publishers are responsible for publishing 28,060 journals. Thirty of these publishers are considered predatory publishers[22].

The official website for each publisher was manually searched for author guidance pertaining to AI-tools broadly, including GAI-based tools. We defined generative-AI (GAI) guidelines as any guidelines mentioning the use of GPTs, LLMs, or GAI. The initial data collection took place between May 19th - 20th 2023 (6 months post-ChatGPT launch). The data collection was completed within a 24-hour period to ensure an accurate snapshot of the available guidance. The data was collected independently by two reviewers (A.P. and B.B.) after training and piloting the data extraction form. Discrepancies were settled by a third reviewer (C.G.) under the supervision of the senior author (G.E.C.). If a publisher's website was in a non-English language, the author guidelines were translated into English using Google Machine Translate, as previously done[23]. If a publisher did not provide any GAI guidance, at least three subsidiary journal websites were evaluated for the existence of shared guidance as a proxy for publisher policy. The data extraction focused on determining the presence of specific author guidance referencing the use of GAI, as well as the date the guidance were released and whether the guidelines referenced any validated reporting criteria for the use of GAI in scientific research.

***Journal Selection and Data Acquisition***

The highest ranked 100 science journals, by H-index, were selected from SCImago.org (https://www.scimagojr.com) on May 4th, 2023 as previously done[24]. The highest ranked journal had a H-index of 1331 and the 100th ranked journal had a H-index of 356.

The official website for each journal was manually searched for guidelines pertaining to AI-tools as described above. The data collection took place between May 19th – 20th 2023. The data collection was completed within a 24-hour period to ensure an accurate snapshot of the available guidance. The data was collected independently by two reviewers (A.P. and B.B.) after training regarding the data extraction. Discrepancies were settled by a third reviewer (C.G.) under the supervision of the senior editor (G.E.C.). If a journal did not provide GAI reporting guidance, the GAI guidance provided by the journal's publisher were used as a proxy only if the author guidance or ethics page directly recommended viewing or linked to the publisher's guidance. Similarly, to the publisher author guidance data collection, the journal author guidance data collection focused on determining the presence of specific author guidance referencing the use of GAI, as well as the date the guidance were released and whether it referenced any validated criteria for the use of GAI in scientific research.

### *Data Presentation*

Descriptive statistics were used to summarise the data, reporting frequencies and percentages for all categorical variables. The denominators for each variable are specified in the results. Charts and tables are used when appropriate to bolster the interpretability of the data. A narrative synthesis was used to describe the study findings.

## RESULTS

All AI guidance identified specifically referred to GAI-based models or the generative ability of AI, in leu of discussing AI use more broadly. Of the largest 100 publishers, 17% have released guidance on GAI. 71% of publishers (n=12) with GAI guidance were in the top 25 publishers. Additionally, 55% of publishers reference membership to the COPE. Of the highest ranked 100 journals, 70% have released GAI guidance. 80% of journals cited membership of COPE. Several of the highest ranked 100 journals shared the same publishers; the most represented publishers included SpringerNature with 19% of the journals in the highest ranked 100 journals, followed by the American Chemical Society with 10%, and Elsevier with 7%.

### *Author Guidance on GAI: Largest 100 Publishers*

Seventeen (17%) publishers had specific AI guidance for authors, twelve (70.6%) of these were in the largest 25 publishers. Seven (41.2%) publishers with GAI guidance also provided a direct link to the COPE position statement on the use of AI in research publications. Among the publisher-specific guidance, 16 (94.1%) publishers had specific guidance for including GAI as an author -- all 16 explicitly stated that GAI may not be listed as an author. There were very few other guidance with explicit prohibitions. Only one (5.9%) other publisher, *Emerald*, specifically had a policy to prohibit the submission of AI-generated images. Three (17.6%) publishers indicated that their guidance only applied to the writing process. Regarding specific GAI tools referenced, 12 (70.6%) publishers mentioned LLMs and 11 (64.7%) explicitly mentioned ChatGPT. All 11 publishers mentioning ChatGPT made no mention any other specific GAI tools.

Concerning the documenting the use of GAI in research, guidance for disclosure generally included a combination of whether to report, where in the manuscript, and/or what details to report. All 17 (100%) publishers with guidance required disclosure in some form, while only 5 (29.4%) specifically highlighted the term “disclose” to describe this process. Twelve (64.7%) publishers provided recommendations on where in the manuscript to include the disclosure, the most common locations being the Methods (n=11, 64.7%), Acknowledgements (n=9, 52.9%), or a similar section (n=4, 23.5%). Additionally, two (11.8%) publishers suggested disclosure in the Cover Letter. Regarding what to disclose, seven (41.2%) publishers had guidance on which details should be included in the disclosure, such as the name, model, and version of the AI tool and the purpose for which AI was used. Only one publisher, *Elsevier*, provided a specific disclosure template to use and advised that it be included in a new, independent section of the manuscript. Finally, 11 (64.7%) publishers stated that the authors are responsible and accountable for the output produced by AI tools. None of the proposed guidance were developed using a formal guideline development process[25]. Details can be found in table 1, figure 1

***Author Guidance on GAI: Highest ranked 100 Journals***

Of the 100 journals, a total of 70 journals had specific GAI guidance for authors. Twenty (28.6%) of these 70 were in the highest ranked 25 journals, while 16 (22.9%) were in the 2nd fourth, 18 (25.7%) in the 3rd fourth, and 16 (22.9%) in the bottom fourth. In addition to individually journal specific guidance, six (8.6%) journals also provided a direct link to the COPE AI-position statement on the use of AI in research publications and two (2.9%), *Lancet* and *Lancet Oncolog*y did not include criteria for the specific use of GAI. Of the 70 journals, four (5.7%) explicitly prohibited any use of GAI tools in the preparation of a manuscript, including *Annals of Internal Medicine, Bioinformatics, Blood,* and *Science*. Other journals explicitly prohibited GAI included *Lancet,* which limited the use of GAI for anything other than improving the "readability and language of the work" and *PLoS ONE's* restriction on using GAI for fabrication or misrepresentation of data. Regarding authorship, 67 (95.7%) journals had specific guidance for including GAI as an author. All of these explicitly stated that AI should not be listed as an author. Fifteen (21.4%) journals indicated that their GAI guidance only applied to the writing process. In regard to specific GAI tools referenced, 36 (51.4%) journals mentioned LLMs and 33 (47.1%) explicitly mentioned ChatGPT. No other specific GAI tools were mentioned in any of the journal policies or guidance.

Guidance for disclosure included a combination of whether, where, and/or what to disclose. Of 70 journals with GAI guidelines, 69 (98.6%) require some type of "reporting," "documenting," or "noting," with *Science* being the only journal without any mention of disclosure. Thirty (42.9%) journals specifically used the term "disclose" to describe this process. Sixty-six (94.3%) journals had guidance on where in the manuscript to include the disclosure, the most common locations being the Methods (n=51, 72.8%), a similar section (n=38, 54.3%), Acknowledgements (n=34, 48.5%), a new section (n=12, 17.1%), and/or the cover letter (n=10, 14.2%). Regarding what to disclose, 31 (44.3%) journals provided recommendations on which details should be included in the disclosure. Ten (14.3%) journals, all *Elsevier* journals, provided a specific disclosure template and advised that it be included in a new, separate section of the manuscript. Finally, 31 (44.3%) journals stated that the authors are responsible and accountable for the output produced by GAI tools. None of the proposed guidelines were developed using any formal guideline development process[25]. Details are reported in table 2, figure 1.

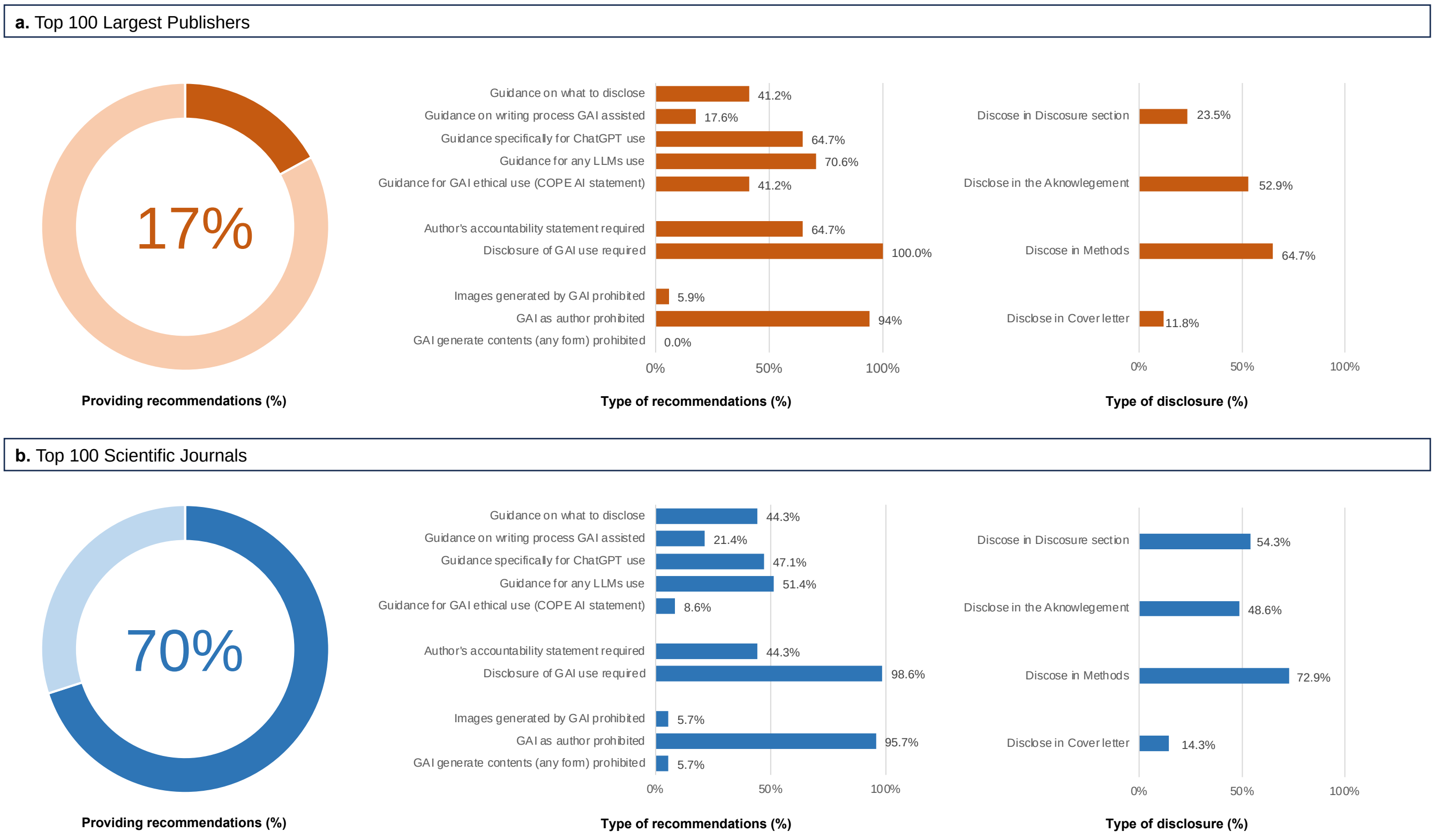

Figure 1. Types of recommendations and types of disclosures recommended in the authors' guidelines from a) the top largest publishers and b) the top-ranked scientific journals.

### *Journal-Publisher Author Guidelines Consistency on GAI*

We found that of all 100 journals, 49 reported GAI guidance or policy on the journal website, of which 11 (22.4%) linked to GAI guidance on their publisher's website. Thirty-five (71.4%) had guidance listed solely on the journal website (i.e., the corresponding publisher did not report GAI guidance and the journals did not link to their publisher). Additionally, 3/49 (6.1%) journals had publishers that also reported guidance, but the journals did not link to the publisher website.

Of the remaining 51 journals that did not report GAI guidance on the journal website, 21 (41.2%) link to the publisher website. Five (9.8%) journals do not link to the publisher website even though the publisher website lists AI guidance. Therefore, 25 (49.0%) journals do not report GAI guidance on the journal website nor the publisher website. Finally, of the 14 journals that provided guidance on the journal website and have publishers that reported GAI guidance, two (14.3%) of these journals have guidance that conflict with those of the publisher (e.g., *American College of Cardiology*).

## DISCUSSION

The present study reveals significant heterogeneity and conflicting author guidance concerning the use of GAI among publishers and journals. We found that less than a quarter of publishers and less than three quarters of the journals currently have some form of guidance in place. All the AI guidance identified specifically referenced GAI models or discussed the generative ability of AI in their statements. As broader AI applications are not discussed, this indicates that journals and publishers developed their policy or author guidance specifically in response to the growing popularity of GAI. There is a notable diversity in the details expressed in the guidance and recommendations posted by each respective publisher and/or journal. None of the proposed guidance were developed using a formal consensus-based guideline development process or supported by evidence.

Out of the 100 publishers, only a few (16%) reported guidance for the use of GAI in research, and most of these publishers were in the top fourth of highest ranked journals. Among publishers, it is notable that *Bentham*, a known "predatory publisher," has created GAI recommendations for authors while other respected publishing houses that catalog thousands of journals have not. Of the publishers that do have GAI guidance, there was little standardization. While most publishers reference their adherence to COPE guidelines, only 7% linked to COPE's position statement on the use of AI[17]. Further, of the publishers that did link to the COPE AI-statement[17], their individual guidance did not always align with the COPE position statement, creating potential confusion. Despite the lack of standardization, there were two trends across different publisher recommendations. One consistently reported principle was that listing GAI as an author is prohibited. The reasons referenced that GAI tools cannot take responsibility for content created, which is a standard requirement throughout author guidelines and consistent with COPE's position statement[17]. Another consistent principle was the requirement to disclose the use of GAI. However, the nature of this disclosure varied significantly. While the majority of publishers specified where in the manuscript to include this disclosure, the suggested locations differed greatly. In some cases, the decision was left to the researcher to disclose in a similar section and several publishers did not propose a location at all. Additionally, less than half of publishers with GAI guidance specified which details to include in the disclosure, such as the name, model, version, source, description of tool, and how the tool was used. Elsevier provided a standardized template that included the name of the tool or service used and the reason for its use. Furthermore, there was variability among publishers in the uses of GAI tools to which the guidelines applied. For instance, while some publisher guidance only pertained to "AI-generated text," others also encompassed the production of images and data analysis. Several of the guidance utilized more generalized verbiage, such as "scholarly contributions," "content creation," and "preparation of a manuscript," introducing another element of confusion. Finally, not all publishers required authors to maintain accountability for the output produced by GAI, setting a precedent for deniability when it comes to ownership of content generated by AI tools.

The majority of the highest ranked 100 journals had guidance on the use of GAI in scientific research. Importantly, many of these journals share the same publishers and are published by larger publishing houses, which also had GAI policy and guidance. In this context, similar themes to the publisher guidance applied to journal guidance. There was great variability across guidance and little standardization. Compared to publishers, an even fewer percentage (6%) of journals linked to the COPE AI position statement[17]. Of the journals that did link to the COPE AI-statement, their journal-specific guidance did not entirely align with the COPE position statement possibly to confusion for authors on which standards to follow. The two most consistent guidance across journals were that GAI cannot be listed as an author and that disclosure of the use of GAI is required in some format. However, there was variability where to disclose, what details to include, and in which format to disclose. Journals published by *Elsevier* had a single set of guidance which requires the use of a standard template. Additionally, there was discordance across journals regarding the uses of GAI that were encapsulated within their AI-tools guidance, with some specifying the writing process, image generation and/or data analysis and collection, while others specified none at all. As identified among publisher policy and guidance, several journals utilized more generalized terms to describe which components of their submission were bound by the GAI guidance. Lastly, less than half of the journals specified that authors are accountable for the output produced by GAI.

Taking the above information into account, we identified sources of heterogeneity in GAI guidance among publishers and journals. In our analysis of different journals and publishers, we noticed inconsistencies in the dissemination of guidance related to GAI. While some journals not only presented their own GAI guidance but also linked directly to the identical guidance provided by their publishers, there were instances where journals issued guidance without providing a link to their publisher's guidance. Conversely, certain journals would solely link to their publisher's guidance without releasing any of their own. This discrepancy results in a lack of centralization of information regarding the use of GAI. Consequently, the responsibility falls onto authors to seek out and understand the available guidance. This setup potentially allows authors to inadvertently misuse GAI tools due to an incomplete understanding of the regulations imposed by their chosen journal or publisher, leading to a potential misuse or misrepresentation of these powerful tools in scientific literature.

In addition to a non-centralized location for information on GAI use, there were also several cases of competing recommendations and guidance (table 3, figure 2). Some journals, such as Journal of the *American College of Cardiology*, had contradicting guidance to what was communicated by their respective publishers. Additionally, several journals and publishers that linked to the COPE AI-position statement endorsed independent guidelines that contradicted COPE's recommendations. These direct incongruities magnify the burden on authors to seek out the "correct" guidance and to decide which standards to follow. We also found heterogeneity in the verbiage used. Guidance frequently used words like "disclose," "report," "describe," "acknowledge," and "document" interchangeably when instructing readers on how to present the use of generative-AI in the manuscript. This can lead to confusion, as these words have discrete definitions - for example, a disclosure of a conflict of interest is not the same as an acknowledgement of a contributor in the context of scientific publishing.

Several journals and publishers do not stipulate that authors are accountable for output produced by GAI. The COPE AI-position statement asserts that authors are "fully responsible" for their work, including any portion produced by AI. This is important because, as some publishers such as *Elsevier* and *SAGE* note in their guidelines, GAI can often produce output that is inaccurate, biased, or misleading[11 13]. GAI are known to "hallucinate" and fabricate unfounded information[11 13]. Additionally, utilizing direct text from generative-AI introduces the risk of plagiarism as the AI may produce duplicated text from its data sources[12 26]. Further, as mentioned in COPE AI-position statement, AI tools are "non-legal entities," that cannot participate in matters of conflict of interest, copyright, and license agreements. Therefore, they do not qualify as authors and cannot take responsibility for any submitted work. In the discussion regarding where to disclose the use of GAI, many journals and publishers recommended disclosure in the acknowledgments section. However, with generative-AI being non-human, lacking agency, and unaccountable for its output, there is reasonable debate regarding whether to GAI should be included in an acknowledgments section alongside collaborators[27].

The elements of heterogeneity discussed above, including the incongruence of GAI guidelines between journals and publishers, misalignment with the COPE position statement, and unclear verbiage surrounding the disclosure of GAI use, could create confusion for authors and reviewers. A lack of clear and standardized recommendations simultaneously place responsibility on authors to seek out guidance, while also diminishing the ethos of these guidance by hindering the ability of authors to appropriately follow them. Standardized recommendations would improve transparency and accountability surrounding GAI use in academia and scientific research. A cross-discipline, global initiative CANGARU (ChatGPT, Generative Artificial Intelligence and Natural Large Language Models for Accountable Reporting and Use Guidelines) is ongoing and results are awaited[21].

This bibliometric analysis represents a snapshot in time six months after the rise in popularity of GAI. As a result, it is likely that GAI guidance have and will continue to change as our understanding of the technology improves and as greater emphasis is placed on creating tailored GAI policies. Additionally, only the largest 100 publishers and highest ranked 100 journals (define by the H-index) were included in this study. It is possible that other publishers or journals have existing GAI guidance of a higher standard that were not represented. Another limitation is that this is a largely qualitative study in which subjectivity of the authors is present. However, this was minimized with a structured system of utilizing multiple reviewers with supervision at several levels from other co-authors.

## CONCLUSION

The landscape of guidance concerning the application of GAI in academic research and scholarly writing exhibits substantial heterogeneity. None of the proposed recommendations were formulated through a structured consensus-based guideline-development process. This scenario highlights an urgent necessity for the establishment of cohesive, cross-disciplinary policies. Such guidance should be carefully crafted in a structured manner, integrating the perspectives of all involved stakeholders. This approach is crucial to counteract the "Babel Tower phenomenon" — the confusion and lack of standardization resulting from individual parties creating their unique regulations.

## DISCLOSURE

ChatGPT3.5 was used for the grammatical check of introduction and discussion paragraphs. The authors validated the output and take full responsibility.

## REFERENCES

1. Allen B, Jr., Seltzer SE, Langlotz CP, et al. A Road Map for Translational Research on Artificial Intelligence in Medical Imaging: From the 2018 National Institutes of Health/RSNA/ACR/The Academy Workshop. *J Am Coll Radiol* 2019;16(9 Pt A):1179-89. doi: 10.1016/j.jacr.2019.04.014 [published Online First: 2019/06/04]
2. Gallego V, Naveiro R, Roca C, et al. AI in drug development: a multidisciplinary perspective. *Mol Divers* 2021;25(3):1461-79. doi: 10.1007/s11030-021-10266-8 [published Online First: 2021/07/13]
3. Rakha EA, Toss M, Shiino S, et al. Current and future applications of artificial intelligence in pathology: a clinical perspective. *J Clin Pathol* 2021;74(7):409-14. doi: 10.1136/jclinpath-2020-206908 [published Online First: 2020/08/09]
4. OpenaI. Introducing ChatGPT OpenAI2022 [updated Nov 30. Available from: https://openai.com/blog/chatgpt.
5. Hu K. ChatGPT sets record for fastest-growing user base - analyst note Reuters2023 [updated Feb 2. Available from: https://www.reuters.com/technology/chatgpt-sets-record-fastest-growing-user-base-analyst-note-2023-02-01/.
6. Newman J. 33 AI tools you can try for free Fast Company2023 [updated Feb 27. Available from: https://www.fastcompany.com/90856183/30-ai-tools-you-can-try-for-free.
7. Pichai S. An important next step on our AI journey: Google; 2023 [updated Feb 6. Available from: https://blog.google/technology/ai/bard-google-ai-search-updates/.
8. Singhal K, Azizi S, Tu T, et al. Large language models encode clinical knowledge. *Nature* 2023 doi: 10.1038/s41586-023-06291-2
9. Graf A, Bernardi RE. ChatGPT in Research: Balancing Ethics, Transparency and Advancement. *Neuroscience* 2023;515:71-73. doi: 10.1016/j.neuroscience.2023.02.008 [published Online First: 2023/02/23]
10. Lecler A, Duron L, Soyer P. Revolutionizing radiology with GPT-based models: Current applications, future possibilities and limitations of ChatGPT. *Diagn Interv Imaging* 2023;104(6):269-74. doi: 10.1016/j.diii.2023.02.003 [published Online First: 2023/03/02]
11. Bhatia G, Kulkarni A. ChatGPT as Co-author: Are researchers impressed or distressed? *Asian J Psychiatr* 2023;84:103564. doi: 10.1016/j.ajp.2023.103564 [published Online First: 2023/03/30]
12. Macdonald C, Adeloye D, Sheikh A, et al. Can ChatGPT draft a research article? An example of population-level vaccine effectiveness analysis. *J Glob Health* 2023;13:01003. doi: 10.7189/jogh.13.01003 [published Online First: 2023/02/18]
13. Ollivier M, Pareek A, Dahmen J, et al. A deeper dive into ChatGPT: history, use and future perspectives for orthopaedic research. *Knee Surg Sports Traumatol Arthrosc* 2023;31(4):1190-92. doi: 10.1007/s00167-023-07372-5 [published Online First: 2023/03/10]
14. Stokel-Walker C. AI bot ChatGPT writes smart essays - should professors worry? *Nature* 2022 doi: 10.1038/d41586-022-04397-7 [published Online First: 2022/12/10]
15. Thorp HH. ChatGPT is fun, but not an author. *Science* 2023;379(6630):313. doi: 10.1126/science.adg7879 [published Online First: 2023/01/27]
16. Sample I. Science journals ban listing of ChatGPT as co-author on papers. *The Guardian* 2023.

17. COPE. Authorship and AI tools: COPE Position Statement: COPE; 2023 [updated Feb 13. Available from: https://publicationethics.org/cope-position-statements/ai-author#:~:text=COPE%20position%20statement&text=COPE%20joins%20organisations%2C%20such%20as,responsibility%20for%20the%20submitted%20work.
18. Flanagin A, Bibbins-Domingo K, Berkwits M, et al. Nonhuman “Authors” and Implications for the Integrity of Scientific Publication and Medical Knowledge. *JAMA* 2023;329(8):637-39. doi: 10.1001/jama.2023.1344
19. Stokel-Walker C. ChatGPT listed as author on research papers: many scientists disapprove. *Nature* 2023;613(7945):620-21. doi: 10.1038/d41586-023-00107-z [published Online First: 2023/01/19]
20. Zielinski C WM, Aggarwal R, Ferris LE, Heinemann M, Lapeña JF, Pai SA, Ing E, Citrome L, Alam M, Voight M, Habibzadeh F, for the WAME Board. Chatbots, Generative AI, and Scholarly Manuscripts. WAME Recommendations on Chatbots and Generative Artificial Intelligence in Relation to Scholarly Publications: WAME, 2023.
21. Cacciamani GE, Collins GS, Gill IS. ChatGPT: standard reporting guidelines for responsible use. *Nature* 2023;618(7964):238. doi: 10.1038/d41586-023-01853-w [published Online First: 2023/06/07]
22. Nishikawa-Pacher A. Who are the 100 largest scientific publishers by journal count? A webscraping approach. *Journal of Documentation* 2022;78(7):450-63.
23. de Vries E, Schoonvelde M, Schumacher G. No Longer Lost in Translation: Evidence that Google Translate Works for Comparative Bag-of-Words Text Applications. *Political Analysis* 2018;26(4):417-30. doi: 10.1017/pan.2018.26 [published Online First: 2018/09/11]
24. Sholklapper TN, Ballon J, Sayegh AS, et al. Bibliometric analysis of academic journal recommendations and requirements for surgical and anesthesiologic adverse events reporting. *Int J Surg* 2023;109(5):1489-96. doi: 10.1097/js9.0000000000000323 [published Online First: 2023/05/03]
25. Moher D, Schulz KF, Simera I, et al. Guidance for Developers of Health Research Reporting Guidelines. *PLOS Medicine* 2010;7(2):e1000217. doi: 10.1371/journal.pmed.1000217
26. Gupta R, Park JB, Bisht C, et al. Expanding Cosmetic Plastic Surgery Research Using ChatGPT. *Aesthet Surg J* 2023 doi: 10.1093/asj/sjad069 [published Online First: 2023/03/22]
27. Paul-Hus A, Desrochers N. Acknowledgements are not just thank you notes: A qualitative analysis of acknowledgements content in scientific articles and reviews published in 2015. *PLoS One* 2019;14(12):e0226727. doi: 10.1371/journal.pone.0226727 [published Online First: 2019/12/20]

*Table 1. Top-100 Publisher Authors Guidelines on Generative Artificial Intelligence*

| Publisher | Number of journals: | Are there any specified guidelines for GAI/GPTs/LLMs? | GAI/GPTs/LLMs Guidelines Reported | Guidelines on how to disclose GAI/GPTs/LLMs |
|---|---|---|---|---|
| **Springer** | 3763 | Yes | Corresponding author(s) should be identified with an asterisk. Large Language Models (LLMs), such as ChatGPT, do not currently satisfy our authorship criteria. Notably an attribution of authorship carries with it accountability for the work, which cannot be effectively applied to LLMs. Use of an LLM should be properly documented in the Methods section (and if a Methods section is not available, in a suitable alternative part) of the manuscript. | Use of an LLM should be properly documented in the Methods section (and if a Methods section is not available, in a suitable alternative part) of the manuscript. |
| **Taylor & Francis** | 2912 | Yes | The use of artificial intelligence (AI) tools in research and writing is an evolving practice. AI-based tools and technologies include but are not limited to large language models (LLMs), generative AI, and chatbots (for example, ChatGPT). Below we restate our guidance on author accountability and responsibilities as it relates to the use of AI tools in content creation. This policy will be iterated as appropriate.Taylor & Francis recognizes the increased use of AI tools in academic research. As the world's leading publisher of human-centered science, we consider that such tools, where used appropriately and responsibly, have the potential to augment research outputs and thus foster progress through knowledge. Authors are accountable for the originality, validity and integrity of the content of their submissions. In choosing to use AI tools, authors are expected to do so responsibly and in accordance with our editorial policies on authorship and principles of publishing ethics.Authorship requires taking accountability for content, consenting to publication via an author publishing agreement, giving contractual assurances about the integrity of the work, among other principles. These are uniquely human responsibilities that cannot be undertaken by AI tools. Therefore, AI tools must not be listed as an author. Authors must, however, acknowledge all sources and contributors included in their work. Where AI tools are used, such use must be acknowledged and documented appropriately. | N/A |
| **Elsevier** | 2674 | Yes | This policy has been triggered by the rise of generative AI and AI-assisted technologies which are expected to increasingly be used by content creators. The policy aims to provide greater transparency and guidance to authors, readers, reviewers, editors and contributors. Elsevier will monitor this development and will adjust or refine this policy when appropriate. Please note the policy only refers to the writing process, and not to the use of AI tools to analyze and draw insights from data as part of the research process.<br><br>Where authors use generative AI and AI-assisted technologies in the writing process, these technologies should only be used to improve readability and language of the work. Applying the technology should be done with human oversight and control and authors should carefully review and edit the result, because AI can generate authoritative-sounding output that can be incorrect, incomplete or biased. The authors are ultimately responsible and accountable for the contents of the work.<br><br>Authors should disclose in their manuscript the use of AI and AI-assisted technologies and a statement will appear in the published work. Declaring the use of these technologies supports transparency and trust between authors, readers, reviewers, editors and contributors and facilitates compliance with the terms of use of the relevant tool or technology.<br><br>Authors should not list AI and AI-assisted technologies as an author or co-author, | Authors must disclose the use of generative AI and AI-assisted technologies in the writing process by adding a statement at the end of their manuscript in the core manuscript file, before the References list. The statement should be placed in a new section entitled ‘Declaration of Generative AI and AI-assisted technologies in the writing process’.<br><br>Statement: During the preparation of this work the author(s) used [NAME TOOL / SERVICE] in order to [REASON]. After using this tool/service, the author(s) reviewed and edited the content as needed and take(s) full responsibility for the content of the publication.<br><br>This declaration does not apply to the use of basic tools for checking grammar, spelling, references etc. If there is nothing to disclose, there is no need to add a statement. |
| | | | nor cite AI as an author. Authorship implies responsibilities and tasks that can only be attributed to and performed by humans. Each (co-) author is accountable for ensuring that questions related to the accuracy or integrity of any part of the work are appropriately investigated and resolved and authorship requires the ability to approve the final version of the work and agree to its submission. Authors are also responsible for ensuring that the work is original, that the stated authors qualify for authorship, and the work does not infringe third party rights, and should familiarize themselves with our Ethics in Publishing policy before they submit. | |
| **Wiley** | 1691 | Yes | Artificial Intelligence Generated Content (AIGC) tools—such as ChatGPT and others based on large language models (LLMs)—cannot be considered capable of initiating an original piece of research without direction by human authors. They also cannot be accountable for a published work or for research design, which is a generally held requirement of authorship (as discussed in the previous section), nor do they have legal standing or the ability to hold or assign copyright. Therefore—in accordance with COPE's position statement on AI tools—these tools cannot fulfill the role of, nor be listed as, an author of an article. If an author has used this kind of tool to develop any portion of a manuscript, its use must be described, transparently and in detail, in the Methods or Acknowledgements section. The author is fully responsible for the accuracy of any information provided by the tool and for correctly referencing any supporting work on which that information depends. Tools that are used to improve spelling, grammar, and general editing are not included in the scope of these guidelines. The final decision about whether use of an AIGC tool is appropriate or permissible in the circumstances of a submitted manuscript or a published article lies with the journal's editor or other party responsible for the publication's editorial policy. | If an author has used this kind of tool to develop any portion of a manuscript, its use must be described, transparently and in detail, in the Methods or Acknowledgements section. |
| **SAGE** | 1208 | yes | Use of Large Language Models and generative AI tools in writing your submission<br><br>Sage recognizes the value of large language models (LLMs) (e.g. ChatGPT) and generative AI as productivity tools that can help authors in preparing their article for submission; to generate initial ideas for a structure, for example, or when summarizing, paraphrasing, language polishing etc. However, it is important to note that all language models have limitations and are unable to replicate human creative and critical thinking. Human intervention with these tools is essential to ensure that content presented is accurate and appropriate to the reader. Sage therefore requires authors to be aware of the limitations of language models and to consider these in any use of LLMs in their submissions:<br><br>Objectivity: Previously published content that contains racist, sexist or other biases can be present in LLM-generated text, and minority viewpoints may not be represented. Use of LLMs has the potential to perpetuate these biases because the information is decontextualized and harder to detect.<br><br>Accuracy: LLMs can 'hallucinate' i.e. generate false content, especially when used outside of their domain or when dealing with complex or ambiguous topics. They can generate content that is linguistically but not scientifically plausible, they can get facts wrong, and they have been shown to generate citations that don't exist. Some LLMs are only trained on content published before | Clearly indicate the use of language models in the manuscript, including which model was used and for what purpose. Please use the methods or acknowledgements section, as appropriate. |
| | | | a particular date and therefore present an incomplete picture.<br><br>Contextual understanding: LLMs cannot apply human understanding to the context of a piece of text, especially when dealing with idiomatic expressions, sarcasm, humour, or metaphorical language. This can lead to errors or misinterpretations in the generated content.<br><br>Training data: LLMs require a large amount of high-quality training data to achieve optimal performance. However, in some domains or languages, such data may not be readily available, limiting the usefulness of the model.<br><br>Guidance for authors<br><br>Authors are required to:<br><br>Clearly indicate the use of language models in the manuscript, including which model was used and for what purpose. Please use the methods or acknowledgements section, as appropriate.<br><br>Verify the accuracy, validity, and appropriateness of the content and any citations generated by language models and correct any errors or inconsistencies.<br><br>Provide a list of sources used to generate content and citations, including those generated by language models. Double-check citations to ensure they are accurate, and are properly referenced.<br><br>Be conscious of the potential for | |
| | | | plagiarism where the LLM may have reproduced substantial text from other sources. Check the original sources to be sure you are not plagiarising someone else's work.<br><br>Acknowledge the limitations of language models in the manuscript, including the potential for bias, errors, and gaps in knowledge.<br><br>Please note that AI bots such as ChatGPT should not be listed as an author on your submission.<br><br>We will take appropriate corrective action where we identify published articles with undisclosed use of such tools.<br><br>Authors should check the guidelines of the journal they are submitting to for any specific policies that may be in place on that journal. | |
| **OMICS** | 705 | No | N/A | N/A |
| **De Gruyter** | 513 | No | N/A | N/A |
| **Oxford University Press** | 500 | Yes | Neither symbolic figures such as Camille Noûs nor natural language processing tools driven by artificial intelligence (AI) such as ChatGPT qualify as authors, and OUP will screen for them in author lists. The use of AI (for example, to help generate content, write code, or analyze data) must be disclosed both in cover letters to editors and in the Methods or Acknowledgements section of manuscripts. | The use of AI (for example, to help generate content, write code, or analyze data) must be disclosed both in cover letters to editors and in the Methods or Acknowledgements section of manuscripts. |
| **InderScience** | 472 | No | N/A | N/A |
| **Brill** | 461 | No | N/A | N/A |
| **Cambridge University Press** | 422 | yes | AI Contributions to Research Content<br>AI use must be declared and clearly explained in publications such as research papers, just as we expect scholars to do with other software, tools and methodologies.<br>AI does not meet the Cambridge requirements for authorship, given the need for accountability. AI and LLM tools may not be listed as an author on any scholarly work published by Cambridge<br>Authors are accountable for the accuracy, integrity and originality of their research papers, including for any use of AI.<br>Any use of AI must not breach Cambridge's plagiarism policy. Scholarly works must be the author's own, and not present others' ideas, data, words or other material without adequate citation and transparent referencing.<br>Please note, individual journals may have more specific requirements or guidelines for upholding this policy. | n/a |
| **Thieme** | 407 | Yes | Thieme aligns itself with the COPE Position Statement on Artificial Intelligence (AI) and Authorship.<br><br>AI tools such as ChatGPT can make scholarly contributions to papers. The use of generative AI tools should be properly documented in in the Acknowledgements or Material and Methods sections. AI tools should not be listed as authors, as they do not fulfil all criteria for authorship: they cannot take responsibility for the integrity and the content of a paper, and they cannot take on legal responsibility.<br><br>Authors are liable for every part of their manuscript, including those parts created with the help of an AI. | The use of generative AI tools should be properly documented in in the Acknowledgements or Material and Methods sections. |
| **Medknow** | 386 | No | N/A | N/A |
| **Emerald** | 377 | Yes | Further to this, and in accordance with COPE's position statement on AI tools, Large Language Models cannot be credited with authorship as they are incapable of conceptualising a research design without human direction and cannot be accountable for the integrity, originality, and validity of the published work.<br><br>Any use of such AI tools for the creation, development, or generation of an Emerald publication must be flagged, clearly and transparently, by the author(s) within the Methods and Acknowledgements (or another appropriate section) of the article, chapter, or case study. The author(s) must describe the content created or modified as well as appropriately cite the name and version of the AI tool used; any additional works drawn on by the AI tool should also be appropriately cited and referenced. Standard tools that are used to improve spelling and grammar are not included within the parameters of this guidance. The Editor and Publisher reserve the right to determine whether the use of an AI tool is permissible in a submitted article, chapter, or case study.<br><br>The submission and publication of images created by AI tools or large-scale generative models is not permitted. | Any use of such AI tools for the creation, development, or generation of an Emerald publication must be flagged, clearly and transparently, by the author(s) within the Methods and Acknowledgements (or another appropriate section) of the article, chapter, or case study. The author(s) must describe the content created or modified as well as appropriately cite the name and version of the AI tool used; any additional works drawn on by the AI tool should also be appropriately cited and referenced. |
| **MDPI** | 376 | Yes | MDPI follows the Committee on Publication Ethics (COPE) position statement when it comes to the use of Artificial Intelligence (AI) and AI-assisted technology in manuscript preparation. Tools such as ChatGPT and other large language models (LLMs) do not meet authorship criteria and thus cannot be listed as authors on manuscripts.<br><br>In situations where AI or AI-assisted tools have been used in the preparation of a manuscript, this must be appropriately declared with sufficient details at submission via the cover letter. Furthermore, authors are required to be transparent about the use of these tools and disclose details of how the AI tool was used within the “Materials and Methods” section, in addition to providing the AI tool’s product details within the “Acknowledgments” section.<br><br>Authors are fully responsible for the originality, validity, and integrity of the content of their manuscript and must ensure that this content complies with all of MDPI’s publication ethics policies. MDPI reserves the right to request further information, and editorial decisions will be made in line with MDPI’s Editorial Process and our Terms and Conditions. | authors are required to be transparent about the use of these tools and disclose details of how the AI tool was used within the “Materials and Methods” section, in addition to providing the AI tool’s product details within the “Acknowledgments” section. |
| **Lippincott, Williams & Wilkins** | 375 | No | N/A | N/A |
| **BioMedCentral** | 306 | Yes | Large Language Models (LLMs), such as ChatGPT, do not currently satisfy our authorship criteria. Notably an attribution of authorship carries with it accountability for the work, which cannot be effectively applied to LLMs. Use of an LLM should be properly documented in the Methods section (and if a Methods section is not available, in a suitable alternative part) of the manuscript. | Use of an LLM should be properly documented in the Methods section (and if a Methods section is not available, in a suitable alternative part) of the manuscript. |
| **IEEE** | 294 | Yes | The use of artificial intelligence (AI)–generated text in an article shall be disclosed in the acknowledgements section of any paper submitted to an IEEE Conference or Periodical. The sections of the paper that use AI-generated text shall have a citation to the AI system used to generate the text. | The use of artificial intelligence (AI)–generated text in an article shall be disclosed in the acknowledgements section of any paper submitted to an IEEE Conference or Periodical. The sections of the paper that use AI-generated text shall have a citation to the AI system used to generate the text. |
| **Science Publishing Group** | 273 | No | N/A | N/A |
| **Philosophy Documentation Center** | 249 | No | N/A | N/A |
| **SCIRP** | 247 | No | N/A | N/A |
| **IRMA** | 244 | No | N/A | N/A |
| **Hindawi** | 243 | No | N/A | N/A |
| **IGI Global** | 238 | No | N/A | N/A |
| **World Scientific** | 204 | yes | World Scientific recognizes that the use of artificial intelligence tools (AI) in academic research and writing is an evolving practice. AI-based tools and technologies include but are not limited to large language models (LLMs), generative AI, and chatbots (for example, ChatGPT).<br>Authors are accountable for the originality and integrity of the content of their manuscript. In choosing to use AI tools, authors are expected to do so responsibly and in accordance with our editorial policies on authorship and principles of publishing ethics.<br><br>Therefore, World Scientific joins COPE to state that AI tools cannot be listed as an author of a paper as they cannot take responsibility for submitted work, and their use should be fully transparent.<br><br>Authors who use AI tools in the writing of a manuscript, production of images or graphical elements of the paper, or in the collection and analysis of data, must be transparent in disclosing in the Materials and Methods (or similar section such as acknowledgement section or introduction) of the paper on how the AI tool was used and which tool was used. The final decision about whether use of an AI generated content tool is appropriate or permissible in a submitted manuscript lies with the journal's editor or other party responsible for the publication's editorial policy. | "how the AI tool was used and which tool was used"<br>"Materials and Methods (or similar section such as acknowledgement section or introduction)" |
| **Austin Publishing Group** | 202 | No | N/A | N/A |
| **Bentham** | 201 | Yes | Bentham Science Publishers recognizes that authors use a variety of tools for preparing articles related to their scientific works, ranging from simple ones to very sophisticated ones.<br><br>According to the COPE (Committee on Publication Ethics) guidelines, "AI tools cannot meet the requirements for authorship as they cannot take responsibility for the submitted work. As non-legal entities, they cannot assert the presence or absence of conflicts of interest nor manage copyright and license agreements".<br><br>The pertinence of such tools may vary and evolve with public opinion, due to which the use of AI-powered language tools has led to a significant debate. These tools may generate useful results, but they can also lead to errors or misleading results; therefore, it is important to know which tools were used for evaluating and interpreting a particular scientific work.<br><br>Considering the above we require that:<br><br>The authors to report any significant use of such tools in their works, such as instruments and software along with text-to-text generative AI consistent with subject standards for methodology.<br>All co-authors should sign a declaration that they take full responsibility for all of its contents, regardless of how the contents were generated. Inappropriate language, plagiarized and biased contents, errors, mistakes, incorrect references, or misleading content generated by AI language tools and the | The authors to report any significant use of such tools in their works, such as instruments and software along with text-to-text generative AI consistent with subject standards for methodology. |
| | | | relevant results reported in scientific works are the full and shared responsibility of all the authors, including co-authors.<br>AI language tools should not be listed as an author; instead, authors should follow clause (1) above. | |
| **Universidade de Sao Paulo** | 200 | No | N/A | N/A |
| **Open Access Pub** | 198 | No | N/A | N/A |
| **Longdom** | 190 | No | N/A | N/A |
| **Universitas Pendidikan Indonesia** | 177 | No | N/A | N/A |
| **Gavin Publishers** | 168 | No | N/A | N/A |
| **Universidad de Buenos Aires** | 168 | No | N/A | N/A |
| **iMedPub** | 163 | No | N/A | N/A |
| **Nauka** | 162 | No | N/A | N/A |
| **Schweizerbart** | 158 | No | N/A | N/A |
| **Fabrizio Serra** | 157 | No | N/A | N/A |
| **Scientific and Academic Publishing** | 149 | No | N/A | N/A |
| **JSciMedCentral** | 147 | No | N/A | N/A |
| **Frontiers** | 138 | No | N/A | N/A |
| **Hans Publishers** | 137 | No | N/A | N/A |
| **Advanced Research Publications** | 135 | No | N/A | N/A |
| **Open Access Text (OAT)** | 134 | No | N/A | N/A |
| **KeAi** | 130 | No | N/A | N/A |
| **eScholarship Publishing** | 128 | No | N/A | N/A |
| **Universidad Nacional Autonoma de Mexico** | 127 | No | N/A | N/A |
| **Intellect Books** | 126 | No | N/A | N/A |
| **Hilaris** | 125 | No | N/A | N/A |
| **Academic Journals** | 125 | No | N/A | N/A |
| **Science and Education Publishing** | 125 | No | N/A | N/A |
| **Universitas Gadjah Mada** | 123 | No | N/A | N/A |
| **Conscientia Beam** | 122 | No | N/A | N/A |
| **Universitas Negeri Semarang** | 120 | No | N/A | N/A |
| **Pleiades** | 119 | No | N/A | N/A |
| **University of Tehran** | 115 | No | N/A | N/A |
| **Sciencedomain International** | 112 | No | N/A | N/A |
| **Karger** | 105 | No | N/A | N/A |
| **Polish Academy of Sciences** | 102 | No | N/A | N/A |
| **IOP Publishing** | 102 | Yes | AI Chatbots or Large Language Models (LLMs) do not meet the minimum authorship criteria set out by IOP Publishing or many other industry authorship guidelines, including WAME and IJCME.  LLMs cannot meet IOPP's requirements for authorship, particularly "Final approval of the version to be published" and "Agreement to be accountable for all aspects of the work in ensuring that questions related to the accuracy or integrity of any part of the work are appropriately investigated and resolved." An author also assumes responsibility for a work, including the need to satisfy any copyright or other legal obligations. The same cannot be applied to LLMs, as they lack the ability or comprehension to assume responsibility for work they have helped create. For example, they cannot understand issues around conflicts of interest, nor do they have the legal personality to sign publishing agreements or licences.<br><br>Authors using LLMs to assist in generating ideas and/or aiding drafting of the paper should declare this fact and provide full transparency of the LLM used (name, version, model, source) within the paper they are submitting. This is in line with IOPP's recommendation to acknowledge any writing assistance. Use of an LLM should be properly documented in the Methods section (and if a Methods section is not available, in the acknowledgment section of the manuscript. Authors using these tools to create any part of their work are requested to check for accuracy and are reminded that they, as named authors | Authors using LLMs to assist in generating ideas and/or aiding drafting of the paper should declare this fact and provide full transparency of the LLM used (name, version, model, source) within the paper they are submitting. This is in line with IOPP's recommendation to acknowledge any writing assistance. Use of an LLM should be properly documented in the Methods section (and if a Methods section is not available, in the acknowledgment section of the manuscript. |
| | | | on the work, take full responsibility for the full content of the work. | |
| **Peertechz Publications** | 101 | No | N/A | N/A |
| **Chinese Academy of Sciences** | 101 | No | N/A | N/A |
| **Mary Ann Liebert** | 101 | No | N/A | N/A |
| **Universidad Nacional de La Plata** | 100 | No | N/A | N/A |
| **John Hopkins University Press** | 100 | No | N/A | N/A |
| **Universitas Airlangga** | 99 | No | N/A | N/A |
| **Universitat de Barcelona** | 98 | No | N/A | N/A |
| **University of Malaya** | 94 | No | N/A | N/A |
| **Universitas Negeri Yogyakarta** | 93 | No | N/A | N/A |
| **Universidade Federal do Espirito Santo** | 93 | No | N/A | N/A |
| **Medcrave** | 93 | No | N/A | N/A |
| **Universidad Nacional de Cordoba** | 92 | No | N/A | N/A |
| **APA** | 92 | No | N/A | N/A |
| **SciTechnol** | 92 | No | N/A | N/A |
| **University of Chicago Press** | 92 | No | N/A | N/A |
| **Universitas Negeri Surabaya** | 91 | No | N/A | N/A |
| **Ubiquity Press** | 91 | No | N/A | N/A |
| **University of Hawaii Press** | 90 | No | N/A | N/A |
| **John Benjamins** | 90 | Yes | All authors are accountable for the originality, validity, and integrity of the paper; for this reason, no Artificial Intelligence qualifies as author. See also the section on 'Artificial Intelligence'. (Addition 22 March 2023) Artificial Intelligence (AI) does not qualify for the role of author (see above) and should not be listed as such. If AI was used in the research or preparation of the paper, this should be declared and explained in the description of the tools or methods used. Any requirements concerning copyright and plagiarism continue to apply. | If AI was used in the research or preparation of the paper, this should be declared and explained in the description of the tools or methods used. |
| **Jagiellonian University Press** | 89 | No | N/A | N/A |
| **Dovepress** | 89 | Yes | Authors must be aware that using AI-based tools and technologies for article content generation, e.g. large language models (LLMs), generative AI, and chatbots (e.g. ChatGPT), is not in line with our authorship criteria.<br><br>All authors are wholly responsible for the originality, validity and integrity of the content of their submissions. Therefore, LLMs and other similar types of tools do not meet the criteria for authorship.<br><br>... Any assistance from AI tools for content generation (e.g. large language models) and other similar types of technical tools which generate article content, must be clearly acknowledged within the article. It is the responsibility of authors to ensure the validity, originality and integrity of their article content. Authors are expected to use these types of tools responsibly and in accordance with our editorial policies on authorship and principles of publishing ethics. | Any assistance from AI tools for content generation (e.g. large language models) and other similar types of technical tools which generate article content, must be clearly acknowledged within the article. |
| **IOS Press** | 89 | No | N/A | N/A |
| **Universidade Federal do Rio Grande do Sul** | 88 | No | N/A | N/A |
| **Universitas Diponegoro** | 87 | No | N/A | N/A |
| **University of Alberta Press** | 87 | No | N/A | N/A |
| **Universidade de Brasilia** | 86 | No | N/A | N/A |
| **Internet Scientific Publications** | 86 | No | N/A | N/A |
| **Adam Mickiewicz University** | 86 | No | N/A | N/A |
| **Penn State University Press** | 84 | No | N/A | N/A |
| **Franco Angeli Edizioni** | 83 | No | N/A | N/A |
| **International Scholars Journals** | 83 | No | N/A | N/A |
| **Annex Publishers** | 82 | No | N/A | N/A |
| **Open Access Journals** | 81 | No | N/A | N/A |
| **Pontificia Universidad Javeriana, Bogota** | 81 | No | N/A | N/A |
| **Herbert Publications** | 81 | No | N/A | N/A |
| **Il Mulino** | 80 | No | N/A | N/A |
| **Medwin Publishers LLC** | 79 | No | N/A | N/A |
| **Premier Publishers** | 78 | No | N/A | N/A |
| **Pulsus Group** | 76 | No | N/A | N/A |
| **Scholarena** | 76 | No | N/A | N/A |
| **Editura Academiei Romane** | 76 | No | N/A | N/A |

*Table 2. Top-100 Journals Authors Guidelines on Generative Artificial Intelligence*

| Journal | Subject | Where are the GAI/GPTs/LLMs guidelines indicated? | Are there any specified guidelines for GAI/GPTs/LLMs? | GAI/GPTs/LLMs Guidelines Reported | Guidelines on how to disclose GAI/GPTs/LLMs |
|---|---|---|---|---|---|
| **Nature** | Multidisciplinary | Journal website | Yes | Large Language Models (LLMs), such as ChatGPT, do not currently satisfy our authorship criteria. Notably an attribution of authorship carries with it accountability for the work, which cannot be effectively applied to LLMs. Use of an LLM should be properly documented in the Methods section (and if a Methods section is not available, in a suitable alternative part) of the manuscript. | Use of an LLM should be properly documented in the Methods section (and if a Methods section is not available, in a suitable alternative part) of the manuscript. |
| **Science** | Art and History, Multidisciplinary | Journal website | Yes | In addition, artificial intelligence tools cannot be authors. Artificial intelligence (AI). Text generated from AI, machine learning, or similar algorithmic tools cannot be used in papers published in Science journals, nor can the accompanying figures, images, or graphics be the products of such tools, without explicit permission from the editors. In addition, an AI program cannot be an author of a Science journal paper. A violation of this policy constitutes scientific misconduct. | N/A |
| **New England Journal of Medicine** | Medicine | N/A | No | N/A | N/A |
| **Cell** | Biochemistry, Genetics and Molecular Biology | both | Yes | The below guidance only refers to the writing process, and not to the use of AI tools to analyze and draw insights from data as part of the research process. Where authors use generative artificial intelligence (AI) and AI-assisted technologies in the writing process, authors should only use these technologies to improve readability and language. Applying the technology should be done with human oversight and control, and authors should carefully review and edit the result, as AI can generate authoritative-sounding output that can be incorrect, incomplete or biased. AI and AI-assisted technologies should not be listed as an author or co-author, or be cited as an author. Authorship implies responsibilities and tasks that can only be attributed to and performed by humans, as outlined in Elsevier's AI policy for authors.<br>Authors should disclose in their manuscript the use of AI and AI-assisted technologies in the writing process by following the instructions below. A statement will appear in the published work. Please note that authors are ultimately responsible and accountable for the contents of the work.<br>Disclosure instructions<br>Authors must disclose the use of generative AI and AI-assisted technologies in the writing process by adding a statement at the end of their manuscript. The statement should be placed in a new section after the 'Declaration of interests' section and the optional 'Inclusion and diversity' section, entitled 'Declaration of Generative AI and AI-assisted technologies in the writing process'.<br>Statement: During the preparation of this work the author(s) used [NAME TOOL / SERVICE] in order to [REASON]. After using this tool/service, the author(s) reviewed and edited the content as needed and take(s) full responsibility for the content of the publication.<br>This declaration does not apply to the use of basic tools for checking grammar, spelling, references etc. If there is nothing to disclose, there is no need to add a statement. | Authors must disclose the use of generative AI and AI-assisted technologies in the writing process by adding a statement at the end of their manuscript. The statement should be placed in a new section after the 'Declaration of interests' section and the optional 'Inclusion and diversity' section, entitled 'Declaration of Generative AI and AI-assisted technologies in the writing process'.<br><br>Statement: During the preparation of this work the author(s) used [NAME TOOL / SERVICE] in order to [REASON]. After using this tool/service, the author(s) reviewed and edited the content as needed and take(s) full responsibility for the content of the publication.<br><br>This declaration does not apply to the use of basic tools for checking grammar, spelling, references etc. If there is nothing to disclose, there is no need to add a statement. |
| **Lancet** | Medicine | both | Yes | Where authors use AI and AI-assisted technologies in the writing process, these technologies should only be used to improve readability and language of the work and not used to replace researcher tasks such as producing scientific insights, analysing and interpreting data, or drawing scientific conclusions. Applying these technologies should only be done with human oversight and control, and authors should carefully review and edit the result because AI can generate authoritative-sounding output that can be incorrect, incomplete, or biased. Authors should not list AI and AI-assisted technologies as an author or co-author, nor cite AI as an author. Authors are ultimately responsible and accountable for the originality, accuracy, and integrity of the work; and should disclose the use of AI and AI-assisted technologies in a statement at the end of the article.<br><br>Clinical trials that report interventions using artificial intelligence must be described according to the CONSORT-AI Extension guidelines and their protocols must be described according to the SPIRIT-AI Extension guidelines | Authors.[..] should disclose the use of AI and AI-assisted technologies in a statement at the end of the article. |
| **Proceedings of the National Academy of Sciences of the United States of America** | Multidisciplinary | Journal website | Yes | Use of artificial intelligence (AI) software, such as ChatGPT, must be noted in the Materials and Methods (or Acknowledgments, if no Materials and Methods section is available) section of the manuscript and may not be listed as an author. | Use of artificial intelligence (AI) software, such as ChatGPT, must be noted in the Materials and Methods (or Acknowledgments, if no Materials and Methods section is available) section of the manuscript and may not be listed as an author. |
| **Chemical Reviews** | Chemistry | Publisher website | Yes | Artificial intelligence (AI) tools do not qualify for authorship. The use of AI tools for text or image generation should be disclosed in the manuscript within the Acknowledgment section with a description of when and how the tools were used. For more substantial use cases or descriptions of AI tool use, authors should provide full details within the Methods or other appropriate section of the manuscript. | The use of AI tools for text or image generation should be disclosed in the manuscript within the Acknowledgment section with a description of when and how the tools were used. For more substantial use cases or descriptions of AI tool use, authors should provide full details within the Methods or other appropriate section of the manuscript. |
| **JAMA - Journal of the American Medical Association** | Medicine | Journal website | Yes | Reproduced and Recreated Material and Image Integrity: The submission and publication of content created by artificial intelligence, language models, machine learning, or similar technologies is discouraged, unless part of formal research design or methods, and is not permitted without clear description of the content that was created and the name of the model or tool, version and extension numbers, and manufacturer. Authors must take responsibility for the integrity of the content generated by these models and tools... Authorship Criteria and Contributions: Nonhuman artificial intelligence, language models, machine learning, or similar technologies do not qualify for authorship. If these models or tools are used to create content or assist with writing or manuscript preparation, authors must take responsibility for the integrity of the content generated by these tools. Authors should report the use of artificial intelligence, language models, machine learning, or similar technologies to create content or assist with writing or editing of manuscripts in the Acknowledgment section or Methods section if this is part of formal research design or methods... Acknowledgement Section: Authors should report the use of artificial intelligence, language models, machine learning, or similar technologies to create content or assist with writing or editing of manuscripts in the Acknowledgment section or the Methods section if this is part of formal research design or methods. This should include a description of the content that was created or edited and the name of the language model or tool, version and extension numbers, and manufacturer. (Note: this does not include basic tools for checking grammar, spelling, references, etc.) | Authors should report the use of artificial intelligence, language models, machine learning, or similar technologies to create content or assist with writing or editing of manuscripts in the Acknowledgment section or the Methods section if this is part of formal research design or methods. This should include a description of the content that was created or edited and the name of the language model or tool, version and extension numbers, and manufacturer. (Note: this does not include basic tools for checking grammar, spelling, references, etc.) |
| **Journal of the American Chemical Society** | Biochemistry, Genetics and Molecular Biology, Chemical Engineering, Chemistry | Publisher website | Yes | Artificial intelligence (AI) tools do not qualify for authorship. The use of AI tools for text or image generation should be disclosed in the manuscript within the Acknowledgment section with a description of when and how the tools were used. For more substantial use cases or descriptions of AI tool use, authors should provide full details within the Methods or other appropriate section of the manuscript. | The use of AI tools for text or image generation should be disclosed in the manuscript within the Acknowledgment section with a description of when and how the tools were used. For more substantial use cases or descriptions of AI tool use, authors should provide full details within the Methods or other appropriate section of the manuscript. |
| **Physical Review Letters** | Physics and Astronomy | N/A | No | N/A | N/A |
| **Circulation** | Medicine | Journal website | Yes | The use of automated assistive writing technologies and tools (commonly referred to as artificial intelligence or machine learning tools) is permitted provided that their use is documented and authors assume responsibility for the content. As with human-generated content, authors are responsible for the accuracy, validity, and originality of computer-generated content. Per ICMJE Authorship Criteria, automated assistive writing technologies do not qualify for authorship as they are unable to provide approval or consent for submission. Per ICMJE recommendations for writing assistance, these tools should be listed in the Acknowledgements; if involved in the research design, the tools should be documented in the Methods. For additional information, see the World Association of Medical Editor recommendations. | Per ICMJE recommendations for writing assistance, these tools should be listed in the Acknowledgements; if involved in the research design, the tools should be documented in the Methods. |
| **Nature Genetics** | Biochemistry, Genetics and Molecular Biology | Journal website | Yes | Large Language Models (LLMs), such as ChatGPT, do not currently satisfy our authorship criteria. Notably an attribution of authorship carries with it accountability for the work, which cannot be effectively applied to LLMs. Use of an LLM should be properly documented in the Methods section (and if a Methods section is not available, in a suitable alternative part) of the manuscript. | Use of an LLM should be properly documented in the Methods section (and if a Methods section is not available, in a suitable alternative part) of the manuscript. |
| **Angewandte Chemie - International Edition** | Chemical Engineering, Chemistry | Publisher website | Yes | Artificial Intelligence Generated Content (AIGC) tools—such as ChatGPT and others based on large language models (LLMs)—cannot be considered capable of initiating an original piece of research without direction by human authors. They also cannot be accountable for a published work or for research design, which is a generally held requirement of authorship (as discussed in the previous section), nor do they have legal standing or the ability to hold or assign copyright. Therefore—in accordance with COPE's position statement on AI tools—these tools cannot fulfill the role of, nor be listed as, an author of an article. If an author has used this kind of tool to develop any portion of a manuscript, its use must be described, transparently and in detail, in the Methods or Acknowledgements section. The author is fully responsible for the accuracy of any information provided by the tool and for correctly referencing any supporting work on which that information depends. Tools that are used to improve spelling, grammar, and general editing are not included in the scope of these guidelines. The final decision about whether use of an AIGC tool is appropriate or permissible in the circumstances of a submitted manuscript or a published article lies with the journal's editor or other party responsible for the publication's editorial policy. | If an author has used this kind of tool to develop any portion of a manuscript, its use must be described, transparently and in detail, in the Methods or Acknowledgements section. |
| **Nucleic Acids Research** | Biochemistry, Genetics and Molecular Biology | Publisher website | Yes | Neither symbolic figures such as Camille Noûs nor natural language processing tools driven by artificial intelligence (AI) such as ChatGPT qualify as authors, and OUP will screen for them in author lists. The use of AI (for example, to help generate content, write code, or analyze data) must be disclosed both in cover letters to editors and in the Methods or Acknowledgements section of manuscripts. | The use of AI (for example, to help generate content, write code, or analyze data) must be disclosed both in cover letters to editors and in the Methods or Acknowledgements section of manuscripts. |
| **Advanced Materials** | Engineering, Materials Science | Publisher website | Yes | Artificial Intelligence Generated Content (AIGC) tools—such as ChatGPT and others based on large language models (LLMs)—cannot be considered capable of initiating an original piece of research without direction by human authors. They also cannot be accountable for a published work or for research design, which is a generally held requirement of authorship (as discussed in the previous section), nor do they have legal standing or the ability to hold or assign copyright. Therefore—in accordance with COPE's position statement on AI tools—these tools cannot fulfill the role of, nor be listed as, an author of an article. If an author has used this kind of tool to develop any portion of a manuscript, its use must be described, transparently and in detail, in the Methods or Acknowledgements section. The author is fully responsible for the accuracy of any information provided by the tool and for correctly referencing any supporting work on which that information depends. Tools that are used to improve spelling, grammar, and general editing are not included in the scope of these guidelines. The final decision about whether use of an AIGC tool is appropriate or permissible in the circumstances of a submitted manuscript or a published article lies with the journal's editor or other party responsible for the publication's editorial policy. | If an author has used this kind of tool to develop any portion of a manuscript, its use must be described, transparently and in detail, in the Methods or Acknowledgements section. |
| **Nature Medicine** | Biochemistry, Genetics and Molecular Biology, Medicine | Journal website | Yes | Large Language Models (LLMs), such as ChatGPT, do not currently satisfy our authorship criteria. Notably an attribution of authorship carries with it accountability for the work, which cannot be effectively applied to LLMs. Use of an LLM should be properly documented in the Methods section (and if a Methods section is not available, in a suitable alternative part) of the manuscript. | Use of an LLM should be properly documented in the Methods section (and if a Methods section is not available, in a suitable alternative part) of the manuscript. |
| **Journal of Clinical Oncology** | Biochemistry, Genetics and Molecular Biology, Medicine | N/A | No | N/A | N/A |
| **Chemical Society Reviews** | Chemistry | Publisher website | Yes | Artificial intelligence (AI) tools, such as ChatGPT or other Large Language Models, cannot be listed as an author on a submitted work. AI tools do not meet the criteria to qualify for authorship, as they are unable to take responsibility for the work, cannot consent to publication nor manage copyright, license or other legal obligations, and are unable to understand issues around conflicts of interest. Any use of AI tools in producing any part of the manuscript must be clearly described in the Experimental or Acknowledgement section. The authors are fully responsible and accountable for the content of their article, including any parts produced by an AI tool. | Any use of AI tools in producing any part of the manuscript must be clearly described in the Experimental or Acknowledgement section. |
| **Journal of Biological Chemistry** | Biochemistry, Genetics and Molecular Biology | N/A | No | N/A | N/A |
| **Nature Materials** | Chemistry, Engineering, Materials Science, Physics and Astronomy | Journal website | Yes | Large Language Models (LLMs), such as ChatGPT, do not currently satisfy our authorship criteria. Notably an attribution of authorship carries with it accountability for the work, which cannot be effectively applied to LLMs. Use of an LLM should be properly documented in the Methods section (and if a Methods section is not available, in a suitable alternative part) of the manuscript. | Use of an LLM should be properly documented in the Methods section (and if a Methods section is not available, in a suitable alternative part) of the manuscript. |
| **Nano Letters** | Chemical Engineering, Chemistry, Engineering, Materials Science, Physics and Astronomy | Publisher website | Yes | Artificial intelligence (AI) tools do not qualify for authorship. The use of AI tools for text or image generation should be disclosed in the manuscript within the Acknowledgment section with a description of when and how the tools were used. For more substantial use cases or descriptions of AI tool use, authors should provide full details within the Methods or other appropriate section of the manuscript. | The use of AI tools for text or image generation should be disclosed in the manuscript within the Acknowledgment section with a description of when and how the tools were used. For more substantial use cases or descriptions of AI tool use, authors should provide full details within the Methods or other appropriate section of the manuscript. |
| **Journal of Clinical Investigation** | Medicine | N/A | No | N/A | N/A |
| **Neuron** | Neuroscience | both | Yes | The below guidance only refers to the writing process, and not to the use of AI tools to analyze and draw insights from data as part of the research process.<br><br>Where authors use generative artificial intelligence (AI) and AI-assisted technologies in the writing process, authors should only use these technologies to improve readability and language. Applying the technology should be done with human oversight and control, and authors should carefully review and edit the result, as AI can generate authoritative-sounding output that can be incorrect, incomplete or biased. AI and AI-assisted technologies should not be listed as an author or co-author, or be cited as an author. Authorship implies responsibilities and tasks that can only be attributed to and performed by humans, as outlined in Elsevier's AI policy for authors.<br><br>Authors should disclose in their manuscript the use of AI and AI-assisted technologies in the writing process by following the instructions below. A statement will appear in the published work. Please note that authors are ultimately responsible and accountable for the contents of the work.<br><br>Disclosure instructions<br>Authors must disclose the use of generative AI and AI-assisted technologies in the writing process by adding a statement at the end of their manuscript. The statement should be placed in a new section after the 'Declaration of interests' section and the optional 'Inclusion and diversity' section, entitled 'Declaration of Generative AI and AI-assisted technologies in the writing process'.<br><br>Statement: During the preparation of this work the author(s) used [NAME TOOL / SERVICE] in order to [REASON]. After using this tool/service, the author(s) reviewed and edited the content as needed and take(s) full responsibility for the content of the | Authors must disclose the use of generative AI and AI-assisted technologies in the writing process by adding a statement at the end of their manuscript. The statement should be placed in a new section after the 'Declaration of interests' section and the optional 'Inclusion and diversity' section, entitled 'Declaration of Generative AI and AI-assisted technologies in the writing process'.<br><br>Statement: During the preparation of this work the author(s) used [NAME TOOL / SERVICE] in order to [REASON]. After using this tool/service, the author(s) reviewed and edited the content as needed and take(s) full responsibility for the content of the publication.<br><br>This declaration does not apply to the use of basic tools for checking grammar, spelling, references etc. If there is nothing to disclose, there is no need to add a statement. |
| | | | | publication.<br><br>This declaration does not apply to the use of basic tools for checking grammar, spelling, references etc. If there is nothing to disclose, there is no need to add a statement. | |
| **Blood** | Biochemistry, Genetics and Molecular Biology, Immunology and Microbiology, Medicine | Journal website | Yes | Machine learning (ML)/artificial intelligence (AI) tools, such as ChatGPT, are not eligible for authorship and may not be listed as an author on submissions to Blood journals. However, research that used ML/AI tools for data acquisition or analysis is eligible for submission. Submissions may include graphic outputs of ML/AI, but the role of ML/AI in creating the graphic must be specified in the legend. Text generated by AI may not be included. | Submissions may include graphic outputs of ML/AI, but the role of ML/AI in creating the graphic must be specified in the legend. |
| **Nature Biotechnology** | Biochemistry, Genetics and Molecular Biology, Chemical Engineering, Chemistry, Immunology and Microbiology | Journal website | Yes | Large Language Models (LLMs), such as ChatGPT, do not currently satisfy our authorship criteria. Notably an attribution of authorship carries with it accountability for the work, which cannot be effectively applied to LLMs. Use of an LLM should be properly documented in the Methods section (and if a Methods section is not available, in a suitable alternative part) of the manuscript. | Use of an LLM should be properly documented in the Methods section (and if a Methods section is not available, in a suitable alternative part) of the manuscript. |
| **Journal of Neuroscience** | Neuroscience | N/A | No | N/A | N/A |
| **Nature Reviews Molecular Cell Biology** | Biochemistry, Genetics and Molecular Biology | Journal website | Yes | Large Language Models (LLMs), such as ChatGPT, do not currently satisfy our authorship criteria. Notably an attribution of authorship carries with it accountability for the work, which cannot be effectively applied to LLMs. Use of an LLM should be properly documented in the Methods section (and if a Methods section is not available, in a suitable alternative part) of the manuscript. | Use of an LLM should be properly documented in the Methods section (and if a Methods section is not available, in a suitable alternative part) of the manuscript. |
| **Nature Reviews Cancer** | Biochemistry, Genetics and Molecular Biology, Medicine | Journal website | Yes | Large Language Models (LLMs), such as ChatGPT, do not currently satisfy our authorship criteria. Notably an attribution of authorship carries with it accountability for the work, which cannot be effectively applied to LLMs. Use of an LLM should be properly documented in the Methods section (and if a Methods section is not available, in a suitable alternative part) of the manuscript. | Use of an LLM should be properly documented in the Methods section (and if a Methods section is not available, in a suitable alternative part) of the manuscript. |
| **Cancer Research** | Biochemistry, Genetics and Molecular Biology, Medicine | N/A | No | N/A | N/A |
| **Physical Review B** | Materials Science, Physics and Astronomy | N/A | No | N/A | N/A |
| **Journal of Experimental Medicine** | Immunology and Microbiology, Medicine | N/A | No | N/A | N/A |
| **BMJ** | Medicine | N/A | No | N/A | N/A |
| **Journal of the American College of Cardiology** | Medicine | both | Yes | Please disclose in the cover letter and in the acknowledgement section (the latter of which is published, if the paper is accepted) if any artificial intelligence (AI) programs (e.g., ChatGPT, or other similar software) contributed to the compilation of the submitted manuscript as well as the nature of the contribution that the tool provided. This could include design, performance, analysis, writing, and reporting of the work.<br><br>Declaration of generative AI in scientific writing<br>The below guidance only refers to the writing process, and not to the use of AI tools to analyse and draw insights from data as part of the research process.<br>Where authors use generative artificial intelligence (AI) and AI-assisted technologies in the writing process, authors should only use these technologies to improve readability and language. Applying the technology should be done with human oversight and control, and authors should carefully review and edit the result, as AI can generate authoritative-sounding output that can be incorrect, incomplete or biased. AI and AI-assisted technologies should not be listed as an author or co-author, or be cited as an author. Authorship implies responsibilities and tasks that can only be attributed to and performed by humans, as outlined in Elsevier's AI policy for authors.<br>Authors should disclose in their manuscript the use of AI and AI-assisted technologies in the writing process by following the instructions below. A statement will appear in the published work. Please note that authors are ultimately responsible and accountable for the contents of the work.<br>Disclosure instructions<br>Authors must disclose the use of generative AI and AI-assisted technologies in the writing process by adding a statement at the end of their manuscript in the core manuscript file, before the References list. The statement should be placed in a new section entitled 'Declaration of Generative AI and AIassisted | Please disclose in the cover letter and in the acknowledgement section (the latter of which is published, if the paper is accepted)... the nature of the contribution that the tool provided<br><br>Authors... should disclose the use of AI and AI-assisted technologies in a statement at the end of the article. |
| | | | | technologies in the writing process'.<br>Statement: During the preparation of this work the author(s) used [NAME TOOL / SERVICE] in order to [REASON]. After using this tool/service, the author(s) reviewed and edited the content as needed and take(s) full responsibility for the content of the publication.<br>This declaration does not apply to the use of basic tools for checking grammar, spelling, references etc. If there is nothing to disclose, there is no need to add a statement. | |
| **Proceedings of the IEEE Computer Society Conference on Computer Vision and Pattern Recognition** | Computer Science | Publisher website | Yes | The use of artificial intelligence (AI)–generated text in an article shall be disclosed in the acknowledgements section of any paper submitted to an IEEE Conference or Periodical. The sections of the paper that use AI-generated text shall have a citation to the AI system used to generate the text. | The use of artificial intelligence (AI)–generated text in an article shall be disclosed in the acknowledgements section of any paper submitted to an IEEE Conference or Periodical. The sections of the paper that use AI-generated text shall have a citation to the AI system used to generate the text. |
| **Genes and Development** | Biochemistry, Genetics and Molecular Biology | N/A | No | N/A | N/A |
| **Applied Physics Letters** | Physics and Astronomy | Publisher website | yes | ChatGPT and similar AI-based large language models should not be listed as an author. As with other instrumentation and software, the use of AI-based large language models such as ChatGPT should be disclosed to editors and reviewers, particularly if they are used to generate significant amounts of text in the manuscript. Authors should provide this information in the appropriate section of their manuscript and to the editor with their submission. | ChatGPT should be disclosed to editors and reviewers, particularly if they are used to generate significant amounts of text in the manuscript. Authors should provide this information in the appropriate section of their manuscript and to the editor with their submission. |
| **Nature Communications** | Biochemistry, Genetics and Molecular Biology, Chemistry, Physics and Astronomy | Journal website | Yes | Large Language Models (LLMs), such as ChatGPT, do not currently satisfy our authorship criteria. Notably an attribution of authorship carries with it accountability for the work, which cannot be effectively applied to LLMs. Use of an LLM should be properly documented in the Methods section (and if a Methods section is not available, in a suitable alternative part) of the manuscript. | Use of an LLM should be properly documented in the Methods section (and if a Methods section is not available, in a suitable alternative part) of the manuscript. |
| **Astrophysical Journal** | Earth and Planetary Sciences, Physics and Astronomy | Publisher website | No | N/A | N/A |
| **Nature Neuroscience** | Neuroscience | Journal website | Yes | Large Language Models (LLMs), such as ChatGPT, do not currently satisfy our authorship criteria. Notably an attribution of authorship carries with it accountability for the work, which cannot be effectively applied to LLMs. Use of an LLM should be properly documented in the Methods section (and if a Methods section is not available, in a suitable alternative part) of the manuscript. | Use of an LLM should be properly documented in the Methods section (and if a Methods section is not available, in a suitable alternative part) of the manuscript. |
| **Environmental Science & Technology** | Chemistry, Environmental Science, Medicine | Publisher website | Yes | Artificial intelligence (AI) tools do not qualify for authorship. The use of AI tools for text or image generation should be disclosed in the manuscript within the Acknowledgment section with a description of when and how the tools were used. For more substantial use cases or descriptions of AI tool use, authors should provide full details within the Methods or other appropriate section of the manuscript. | The use of AI tools for text or image generation should be disclosed in the manuscript within the Acknowledgment section with a description of when and how the tools were used. For more substantial use cases or descriptions of AI tool use, authors should provide full details within the Methods or other appropriate section of the manuscript. |
| **Nature Reviews Neuroscience** | Neuroscience | Journal website | Yes | Large Language Models (LLMs), such as ChatGPT, do not currently satisfy our authorship criteria. Notably an attribution of authorship carries with it accountability for the work, which cannot be effectively applied to LLMs. Use of an LLM should be properly documented in the Methods section (and if a Methods section is not available, in a suitable alternative part) of the manuscript. | Use of an LLM should be properly documented in the Methods section (and if a Methods section is not available, in a suitable alternative part) of the manuscript. |
| **ACS Nano** | Engineering, Materials Science, Physics and Astronomy | Publisher website | Yes | Artificial intelligence (AI) tools do not qualify for authorship. The use of AI tools for text or image generation should be disclosed in the manuscript within the Acknowledgment section with a description of when and how the tools were used. For more substantial use cases or descriptions of AI tool use, authors should provide full details within the Methods or other appropriate section of the manuscript. | The use of AI tools for text or image generation should be disclosed in the manuscript within the Acknowledgment section with a description of when and how the tools were used. For more substantial use cases or descriptions of AI tool use, authors should provide full details within the Methods or other appropriate section of the manuscript. |
| **Lecture Notes in Computer Science** | Computer Science, Mathematics | Publisher website | No | N/A | N/A |
| **Nature Reviews Immunology** | Immunology and Microbiology, Medicine | Journal website | Yes | Large Language Models (LLMs), such as ChatGPT, do not currently satisfy our authorship criteria. Notably an attribution of authorship carries with it accountability for the work, which cannot be effectively applied to LLMs. Use of an LLM should be properly documented in the Methods section (and if a Methods section is not available, in a suitable alternative part) of the manuscript. | Use of an LLM should be properly documented in the Methods section (and if a Methods section is not available, in a suitable alternative part) of the manuscript. |
| **Bioinformatics** | Biochemistry, Genetics and Molecular Biology, Computer Science, Mathematics | both | yes | The International Society for Computational Biology (ISCB) has created an acceptable use policy for large language models (LLMs), which the journal follows. It is likely that these guidelines will be subject to change in the future as the development of these models continues to change.<br><br>Common Acceptable Uses:<br><br>As an aid to correct written text (spell checkers, grammar checkers)<br>As an aid to language translation, however, the researcher is responsible for the accuracy of the final text<br>As an algorithmic technique for research study<br>As an evaluation technique (to assist in finding inconsistencies or other anomalies)<br>It is permissible to include LLM generated text snippets as examples in research papers where appropriate, but these MUST be clearly labelled, and their use explained.<br>Assist in code writing, however, the researcher is responsible for the correct code<br>Create documentation for code, however, the researcher is responsible for the correct documentation<br>Any acceptable use of LLMs or related technologies to produce, or help to produce, part of the text, figures or other contents of the paper should be explicitly declared and documented with sufficient details in the supplementary materials.<br><br>Unacceptable Uses:<br><br>It is not acceptable to use LLMs or related technologies to draft papers (including but not limited to text, figures, tables, and references) from a prompt text. In essence, papers must be written by researchers.<br>LLMs cannot be listed as authors as they would not fulfil the requirements of authorship as laid out in the ICMJE guidelines. | (cover letters to editors and in the Methods or Acknowledgements), sufficient details in the supplementary materials.<br>It is permissible to include LLM generated text snippets as examples in research papers where appropriate, but these MUST be clearly labelled, and their use explained. |
| | | | | Natural language processing tools driven by artificial intelligence (AI) do not qualify as authors, and the Journal will screen for them in author lists. The use of AI (for example, to help generate content, write code, or process data) should be disclosed both in cover letters to editors and in the Methods or Acknowledgements section of manuscripts. Please see the COPE position statement on Authorship and AI for more details.<br><br>If your usage of LLMs is not covered by any of these use cases, then please contact the Editor of the journal or Editorial Office. | |
| **Gastroenterology** | Medicine | N/A | No | N/A | N/A |
| **Accounts of Chemical Research** | Chemistry, Medicine | Publisher website | Yes | Artificial intelligence (AI) tools do not qualify for authorship. The use of AI tools for text or image generation should be disclosed in the manuscript within the Acknowledgment section with a description of when and how the tools were used. For more substantial use cases or descriptions of AI tool use, authors should provide full details within the Methods or other appropriate section of the manuscript. | The use of AI tools for text or image generation should be disclosed in the manuscript within the Acknowledgment section with a description of when and how the tools were used. For more substantial use cases or descriptions of AI tool use, authors should provide full details within the Methods or other appropriate section of the manuscript. |
| **Immunity** | Immunology and Microbiology, Medicine | both | Yes | The below guidance only refers to the writing process, and not to the use of AI tools to analyze and draw insights from data as part of the research process.<br><br>Where authors use generative artificial intelligence (AI) and AI-assisted technologies in the writing process, authors should only use these technologies to improve readability and language. Applying the technology should be done with human oversight and control, and authors should carefully review and edit the result, as AI can generate authoritative-sounding output that can be incorrect, incomplete or biased. AI and AI-assisted technologies should not be listed as an author or co-author, or be cited as an author. Authorship implies responsibilities and tasks that can only be attributed to and performed by humans, as outlined in Elsevier's AI policy for authors.<br><br>Authors should disclose in their manuscript the use of AI and AI-assisted technologies in the writing process by following the instructions below. A statement will appear in the published work. Please note that authors are ultimately responsible and accountable for the contents of the work.<br><br>Disclosure instructions<br>Authors must disclose the use of generative AI and AI-assisted technologies in the writing process by adding a statement at the end of their manuscript. The statement should be placed in a new section after the 'Declaration of interests' section and the optional 'Inclusion and diversity' section, entitled 'Declaration of Generative AI and AI-assisted technologies in the writing process'.<br><br>Statement: During the preparation of this work the author(s) used [NAME TOOL / SERVICE] in order to [REASON]. After using this tool/service, the author(s) reviewed and edited the content as needed and take(s) full responsibility for the content of the | Authors must disclose the use of generative AI and AI-assisted technologies in the writing process by adding a statement at the end of their manuscript. The statement should be placed in a new section after the 'Declaration of interests' section and the optional 'Inclusion and diversity' section, entitled 'Declaration of Generative AI and AI-assisted technologies in the writing process'.<br><br>Statement: During the preparation of this work the author(s) used [NAME TOOL / SERVICE] in order to [REASON]. After using this tool/service, the author(s) reviewed and edited the content as needed and take(s) full responsibility for the content of the publication.<br><br>This declaration does not apply to the use of basic tools for checking grammar, spelling, references etc. If there is nothing to disclose, there is no need to add a statement. |
| | | | | publication.<br><br>This declaration does not apply to the use of basic tools for checking grammar, spelling, references etc. If there is nothing to disclose, there is no need to add a statement. | |
| **Molecular Cell** | Biochemistry, Genetics and Molecular Biology | both | Yes | The below guidance only refers to the writing process, and not to the use of AI tools to analyze and draw insights from data as part of the research process.<br><br>Where authors use generative artificial intelligence (AI) and AI-assisted technologies in the writing process, authors should only use these technologies to improve readability and language. Applying the technology should be done with human oversight and control, and authors should carefully review and edit the result, as AI can generate authoritative-sounding output that can be incorrect, incomplete or biased. AI and AI-assisted technologies should not be listed as an author or co-author, or be cited as an author. Authorship implies responsibilities and tasks that can only be attributed to and performed by humans, as outlined in Elsevier's AI policy for authors.<br><br>Authors should disclose in their manuscript the use of AI and AI-assisted technologies in the writing process by following the instructions below. A statement will appear in the published work. Please note that authors are ultimately responsible and accountable for the contents of the work.<br><br>Disclosure instructions<br>Authors must disclose the use of generative AI and AI-assisted technologies in the writing process by adding a statement at the end of their manuscript. The statement should be placed in a new section after the 'Declaration of interests' section and the optional 'Inclusion and diversity' section, entitled 'Declaration of Generative AI and AI-assisted technologies in the writing process'.<br><br>Statement: During the preparation of this work the author(s) used [NAME TOOL / SERVICE] in order to [REASON]. After using this tool/service, the author(s) reviewed and edited the content as needed and take(s) full responsibility for the content of the | Authors must disclose the use of generative AI and AI-assisted technologies in the writing process by adding a statement at the end of their manuscript. The statement should be placed in a new section after the 'Declaration of interests' section and the optional 'Inclusion and diversity' section, entitled 'Declaration of Generative AI and AI-assisted technologies in the writing process'.<br><br>Statement: During the preparation of this work the author(s) used [NAME TOOL / SERVICE] in order to [REASON]. After using this tool/service, the author(s) reviewed and edited the content as needed and take(s) full responsibility for the content of the publication.<br><br>This declaration does not apply to the use of basic tools for checking grammar, spelling, references etc. If there is nothing to disclose, there is no need to add a statement. |
| | | | | publication.<br><br>This declaration does not apply to the use of basic tools for checking grammar, spelling, references etc. If there is nothing to disclose, there is no need to add a statement. | |
| **Nature Immunology** | Immunology and Microbiology, Medicine | Journal website | Yes | Large Language Models (LLMs), such as ChatGPT, do not currently satisfy our authorship criteria. Notably an attribution of authorship carries with it accountability for the work, which cannot be effectively applied to LLMs. Use of an LLM should be properly documented in the Methods section (and if a Methods section is not available, in a suitable alternative part) of the manuscript. | Use of an LLM should be properly documented in the Methods section (and if a Methods section is not available, in a suitable alternative part) of the manuscript. |
| **Annals of Internal Medicine** | Medicine | Journal website | Yes | At submission, Annals requires authors to attest whether they used Artificial Intelligence (AI)-assisted technologies (such as Large Language Models (LLMs), chatbots or image creators) in the production of submitted work. Authors who use such technology should describe, in both the cover letter and the submitted work, how they used it. Chatbots (such as ChatGPT) should not be listed as authors because they cannot be responsible for the accuracy, integrity, and originality of the work, and these responsibilities are required for authorship https://www.icmje.org/recommendations/. Therefore, human authors are responsible for any submitted material that included the use of AI-assisted technologies.<br><br>Corresponding author(s) should be identified with an asterisk. Large Language Models (LLMs), such as ChatGPT, do not currently satisfy our authorship criteria. Notably an attribution of authorship carries with it accountability for the work, which cannot be effectively applied to LLMs. Use of an LLM should be properly documented in the Methods section (and if a Methods section is not available, in a suitable alternative part) of the manuscript.<br><br>Annals discourages the use of artificial intelligence to assist in the review of manuscripts. Under no circumstance should reviewers upload a manuscript, associated files, a description of the manuscript, or your reviewer comments to any Artificial Intelligence tools such as Chat GPT as doing so would violate the confidentiality agreement between the authors and the journal. The reviewer will receive a copy of our decision letter to the author with the other reviewers' comments. These are also confidential. At submission, Annals requires authors to attest whether they used Artificial Intelligence (AI)-assisted technologies (such as Large Language Models (LLMs), chatbots or image creators) in the production of submitted work. Authors who use such technology | Authors who use such technology should describe, in both the cover letter and the submitted work, how they used it. |
| | | | | should describe, in both the cover letter and the submitted work, how they used it. Chatbots (such as ChatGPT) should not be listed as authors because they cannot be responsible for the accuracy, integrity, and originality of the work, and these responsibilities are required for authorship https://www.icmje.org/recommendations/. Therefore, human authors are responsible for any submitted material that included the use of AI-assisted technologies. | |
| **Biomaterials** | Biochemistry, Genetics and Molecular Biology, Chemical Engineering, Engineering, Materials Science | both | Yes | The below guidance only refers to the writing process, and not to the use of AI tools to analyze and draw insights from data as part of the research process.<br><br>Where authors use generative artificial intelligence (AI) and AI-assisted technologies in the writing process, authors should only use these technologies to improve readability and language. Applying the technology should be done with human oversight and control, and authors should carefully review and edit the result, as AI can generate authoritative-sounding output that can be incorrect, incomplete or biased. AI and AI-assisted technologies should not be listed as an author or co-author, or be cited as an author. Authorship implies responsibilities and tasks that can only be attributed to and performed by humans, as outlined in Elsevier's AI policy for authors.<br><br>Authors should disclose in their manuscript the use of AI and AI-assisted technologies in the writing process by following the instructions below. A statement will appear in the published work. Please note that authors are ultimately responsible and accountable for the contents of the work.<br><br>Disclosure instructions<br>Authors must disclose the use of generative AI and AI-assisted technologies in the writing process by adding a statement at the end of their manuscript. The statement should be placed in a new section after the 'Declaration of interests' section and the optional 'Inclusion and diversity' section, entitled 'Declaration of Generative AI and AI-assisted technologies in the writing process'.<br><br>Statement: During the preparation of this work the author(s) used [NAME TOOL / SERVICE] in order to [REASON]. After using this tool/service, the author(s) reviewed and edited the content as needed and take(s) full responsibility for the content of the | Authors must disclose the use of generative AI and AI-assisted technologies in the writing process by adding a statement at the end of their manuscript. The statement should be placed in a new section after the 'Declaration of interests' section and the optional 'Inclusion and diversity' section, entitled 'Declaration of Generative AI and AI-assisted technologies in the writing process'.<br><br>Statement: During the preparation of this work the author(s) used [NAME TOOL / SERVICE] in order to [REASON]. After using this tool/service, the author(s) reviewed and edited the content as needed and take(s) full responsibility for the content of the publication.<br><br>This declaration does not apply to the use of basic tools for checking grammar, spelling, references etc. If there is nothing to disclose, there is no need to add a statement. |
| | | | | publication.<br><br>This declaration does not apply to the use of basic tools for checking grammar, spelling, references etc. If there is nothing to disclose, there is no need to add a statement. | |
| **EMBO Journal** | Biochemistry, Genetics and Molecular Biology, Immunology and Microbiology, Medicine, Neuroscience | N/A | No | N/A | N/A |
| **Journal of Personality and Social Psychology** | Psychology, Social Sciences | N/A | No | N/A | N/A |
| **Journal of Physical Chemistry B** | Chemistry, Material Science, Medicine | Publisher website | Yes | Artificial intelligence (AI) tools do not qualify for authorship. The use of AI tools for text or image generation should be disclosed in the manuscript within the Acknowledgment section with a description of when and how the tools were used. For more substantial use cases or descriptions of AI tool use, authors should provide full details within the Methods or other appropriate section of the manuscript. | The use of AI tools for text or image generation should be disclosed in the manuscript within the Acknowledgment section with a description of when and how the tools were used. For more substantial use cases or descriptions of AI tool use, authors should provide full details within the Methods or other appropriate section of the manuscript. |
| **Chemistry of Materials** | Chemical Engineering, Chemistry, Materials Science | Publisher website | Yes | Artificial intelligence (AI) tools do not qualify for authorship. The use of AI tools for text or image generation should be disclosed in the manuscript within the Acknowledgment section with a description of when and how the tools were used. For more substantial use cases or descriptions of AI tool use, authors should provide full details within the Methods or other appropriate section of the manuscript. | The use of AI tools for text or image generation should be disclosed in the manuscript within the Acknowledgment section with a description of when and how the tools were used. For more substantial use cases or descriptions of AI tool use, authors should provide full details within the Methods or other appropriate section of the manuscript. |
| **Energy and Environmental Science** | Energy, Environmental Science | Publisher website | Yes | Artificial intelligence (AI) tools, such as ChatGPT or other Large Language Models, cannot be listed as an author on a submitted work. AI tools do not meet the criteria to qualify for authorship, as they are unable to take responsibility for the work, cannot consent to publication nor manage copyright, licence or other legal obligations, and are unable to understand issues around conflicts of interest. Any use of AI tools in producing any part of the manuscript must be clearly described in the Experimental or Acknowledgement section. The authors are fully responsible and accountable for the content of their article, including any parts produced by an AI tool. | Any use of AI tools in producing any part of the manuscript must be clearly described in the Experimental or Acknowledgement section. |
| **Nature Reviews Genetics** | Biochemistry, Genetics and Molecular Biology, Medicine | Journal website | Yes | Large Language Models (LLMs), such as ChatGPT, do not currently satisfy our authorship criteria. Notably an attribution of authorship carries with it accountability for the work, which cannot be effectively applied to LLMs. Use of an LLM should be properly documented in the Methods section (and if a Methods section is not available, in a suitable alternative part) of the manuscript. | Use of an LLM should be properly documented in the Methods section (and if a Methods section is not available, in a suitable alternative part) of the manuscript. |
| **Journal of Cell Biology** | Biochemistry, Genetics and Molecular Biology, Medicine | N/A | No | N/A | N/A |
| **American Journal of Respiratory and Critical Care Medicine** | Medicine | Journal website | Yes | No large language model (LLM)-driven chatbots, including ChatGPT, will be accepted as a credited author on a research paper. All author attributions must demonstrate accountability for the work, and AI tools cannot take such responsibility. Researchers using LLM tools should document this use in the methods or acknowledgments sections. If a paper does not include these sections, the introduction or another appropriate section can be used to document the use of the LLM. For more information, please see the below link from COPE.<br><br>https://publicationethics.org/cope-position-statements/ai-author | Researchers using LLM tools should document this use in the methods or acknowledgments sections. If a paper does not include these sections, the introduction or another appropriate section can be used to document the use of the LLM. |
| **PLoS ONE** | Multidisciplinary | Journal website | Yes | PLOS expects that articles should report the listed authors' own work and ideas. Any contributions made by other sources must be clearly and correctly attributed.<br><br>Contributions by artificial intelligence (AI) tools and technologies to a study or to an article's contents must be clearly reported in a dedicated section of the Methods, or in the Acknowledgements section for article types lacking a Methods section. This section should include the name(s) of any tools used, a description of how the authors used the tool(s) and evaluated the validity of the tool's outputs, and a clear statement of which aspects of the study, article contents, data, or supporting files were affected/generated by AI tool usage.<br><br>In cases where Large Language Model (LLM) AI tools or technologies contribute to generating text content for a PLOS submission, the article's authors are responsible for ensuring that:<br><br>the content is accurate and valid,<br>there are no concerns about potential plagiarism,<br>all relevant sources are cited,<br>all statements in the article reporting hypotheses, interpretations, results, conclusions, limitations, and implications of the study represent the authors' own ideas.<br>The use of AI tools and technologies to fabricate or otherwise misrepresent primary research data is unacceptable.<br><br>Noncompliance with any aspect of this policy will be considered misrepresentation of methods, contributions, and/or results. If concerns arise about noncompliance with this policy, PLOS may notify the authors' institution(s) and the journal may reject (pre-publication), retract (post-publication), or publish an editorial notice on the article... PLOS does not allow artificial intelligence (AI) tools and technologies to be listed as authors. If AI tools were | Contributions by artificial intelligence (AI) tools and technologies to a study or to an article's contents must be clearly reported in a dedicated section of the Methods, or in the Acknowledgements section for article types lacking a Methods section. This section should include the name(s) of any tools used, a description of how the authors used the tool(s) and evaluated the validity of the tool's outputs, and a clear statement of which aspects of the study, article contents, data, or supporting files were affected/generated by AI tool usage. |
| | | | | used in conducting the study or preparing the manuscript, their usage must be disclosed transparently in the Methods section (or the Acknowledgements for article types lacking a Methods section) and the article must clearly report which content was affected. See our Artificial Intelligence Tools and Technologies policy for more information about our requirements. | |
| **Journal of Immunology** | Immunology and Microbiology, Medicine | N/A | No | N/A | N/A |
| **Diabetes Care** | Medicine, Nursing | Journal website | Yes | Authorship: Lastly, ADA has adopted and modified JAMA's instructions for authors on the role of artificial intelligence and machine learning in creating content or assisting with writing or manuscript preparation. First, nonhuman artificial intelligence, language models, machine learning, or similar technologies do not qualify for authorship.<br><br>Second, if these models or tools are used to create content or assist with writing or manuscript preparation, authors must take responsibility for the integrity of the content generated by these tools.<br><br>Third, authors should report the use of artificial intelligence, language models, machine learning, or similar technologies to create content or assist with writing or editing of manuscripts in the Acknowledgments section or the Methods section if this is part of formal research design or methods, as well as in the comments to the editors at the time of submission.<br><br>This should include a description of the content that was created or edited and the name of the language model or tool, version and extension numbers, manufacturer, and (where relevant) the query or prompt to create or assist with the development of content. (Note: this does not include basic tools for checking grammar, spelling, references, etc.)...<br>Figures: ADA has adopted and modified JAMA's instructions for authors on the role of artificial intelligence and machine learning in reproducing and re-creating material for publication. In particular, the submission and publication of content created by artificial intelligence, language models, machine learning, or similar technologies is discouraged, unless part of formal research design or methods, and is not permitted without clear description of the content that was created and the name of the model or tool, version and extension numbers, manufacturer, and (where relevant) the query or prompt to create or assist with the development of | Authors should report the use of artificial intelligence, language models, machine learning, or similar technologies to create content or assist with writing or editing of manuscripts in the Acknowledgment section or the Methods section if this is part of formal research design or methods. This should include a description of the content that was created or edited and the name of the language model or tool, version and extension numbers, and manufacturer. (Note: this does not include basic tools for checking grammar, spelling, references, etc.) |
| | | | | content. Authors must take responsibility for the integrity of the content generated by these models and tools.<br>...Image Development: ADA has adopted and modified JAMA's instructions for authors on the role of artificial intelligence and machine learning on the development of images presented for publication. In particular, the submission and publication of images created by artificial intelligence, machine learning tools, or similar technologies is discouraged, unless part of formal research design or methods, and is not permitted without clear description of the content that was created and the name of the model or tool, version and extension numbers, manufacturer, and (where relevant) the query or prompt to create or assist with the development of content. Authors must take responsibility for the integrity of the content generated by these models and tools. | |
| **NeuroImage** | Neuroscience | both | Yes | The below guidance only refers to the writing process, and not to the use of AI tools to analyze and draw insights from data as part of the research process.<br><br>Where authors use generative artificial intelligence (AI) and AI-assisted technologies in the writing process, authors should only use these technologies to improve readability and language. Applying the technology should be done with human oversight and control, and authors should carefully review and edit the result, as AI can generate authoritative-sounding output that can be incorrect, incomplete or biased. AI and AI-assisted technologies should not be listed as an author or co-author, or be cited as an author. Authorship implies responsibilities and tasks that can only be attributed to and performed by humans, as outlined in Elsevier's AI policy for authors.<br><br>Authors should disclose in their manuscript the use of AI and AI-assisted technologies in the writing process by following the instructions below. A statement will appear in the published work. Please note that authors are ultimately responsible and accountable for the contents of the work.<br><br>Disclosure instructions<br>Authors must disclose the use of generative AI and AI-assisted technologies in the writing process by adding a statement at the end of their manuscript. The statement should be placed in a new section after the 'Declaration of interests' section and the optional 'Inclusion and diversity' section, entitled 'Declaration of Generative AI and AI-assisted technologies in the writing process'.<br><br>Statement: During the preparation of this work the author(s) used [NAME TOOL / SERVICE] in order to [REASON]. After using this tool/service, the author(s) reviewed and edited the content as needed and take(s) full responsibility for the content of the | Authors must disclose the use of generative AI and AI-assisted technologies in the writing process by adding a statement at the end of their manuscript. The statement should be placed in a new section after the 'Declaration of interests' section and the optional 'Inclusion and diversity' section, entitled 'Declaration of Generative AI and AI-assisted technologies in the writing process'.<br><br>Statement: During the preparation of this work the author(s) used [NAME TOOL / SERVICE] in order to [REASON]. After using this tool/service, the author(s) reviewed and edited the content as needed and take(s) full responsibility for the content of the publication.<br><br>This declaration does not apply to the use of basic tools for checking grammar, spelling, references etc. If there is nothing to disclose, there is no need to add a statement. |
| | | | | publication.<br><br>This declaration does not apply to the use of basic tools for checking grammar, spelling, references etc. If there is nothing to disclose, there is no need to add a statement. | |
| **IEEE Transactions on Pattern Analysis and Machine Intelligence** | Computer Science, Mathematics | Publisher website | Yes | The use of artificial intelligence (AI)–generated text in an article shall be disclosed in the acknowledgements section of any paper submitted to an IEEE Conference or Periodical. The sections of the paper that use AI-generated text shall have a citation to the AI system used to generate the text. | The use of artificial intelligence (AI)–generated text in an article shall be disclosed in the acknowledgements section of any paper submitted to an IEEE Conference or Periodical. The sections of the paper that use AI-generated text shall have a citation to the AI system used to generate the text. |
| **Nature Cell Biology** | Biochemistry, Genetics and Molecular Biology | Journal website | Yes | Large Language Models (LLMs), such as ChatGPT, do not currently satisfy our authorship criteria. Notably an attribution of authorship carries with it accountability for the work, which cannot be effectively applied to LLMs. Use of an LLM should be properly documented in the Methods section (and if a Methods section is not available, in a suitable alternative part) of the manuscript. | Use of an LLM should be properly documented in the Methods section (and if a Methods section is not available, in a suitable alternative part) of the manuscript. |
| **Neurology** | Medicine | N/A | No | N/A | N/A |
| **Nature Nanotechnology** | Chemical Engineering, Engineering, Materials Science, Physics and Astronomy | Journal website | Yes | Large Language Models (LLMs), such as ChatGPT, do not currently satisfy our authorship criteria. Notably an attribution of authorship carries with it accountability for the work, which cannot be effectively applied to LLMs. Use of an LLM should be properly documented in the Methods section (and if a Methods section is not available, in a suitable alternative part) of the manuscript. | Use of an LLM should be properly documented in the Methods section (and if a Methods section is not available, in a suitable alternative part) of the manuscript. |
| **JAMA Psychiatry** | Medicine | Journal website | Yes | Reproduced and Recreated Material and Image Integrity: The submission and publication of content created by artificial intelligence, language models, machine learning, or similar technologies is discouraged, unless part of formal research design or methods, and is not permitted without clear description of the content that was created and the name of the model or tool, version and extension numbers, and manufacturer. Authors must take responsibility for the integrity of the content generated by these models and tools... Authorship Criteria and Contributions: Nonhuman artificial intelligence, language models, machine learning, or similar technologies do not qualify for authorship. If these models or tools are used to create content or assist with writing or manuscript preparation, authors must take responsibility for the integrity of the content generated by these tools. Authors should report the use of artificial intelligence, language models, machine learning, or similar technologies to create content or assist with writing or editing of manuscripts in the Acknowledgment section or Methods section if this is part of formal research design or methods... Acknowledgement Section: Authors should report the use of artificial intelligence, language models, machine learning, or similar technologies to create content or assist with writing or editing of manuscripts in the Acknowledgment section or the Methods section if this is part of formal research design or methods. This should include a description of the content that was created or edited and the name of the language model or tool, version and extension numbers, and manufacturer. (Note: this does not include basic tools for checking grammar, spelling, references, etc.) | Authors should report the use of artificial intelligence, language models, machine learning, or similar technologies to create content or assist with writing or editing of manuscripts in the Acknowledgment section or the Methods section if this is part of formal research design or methods. This should include a description of the content that was created or edited and the name of the language model or tool, version and extension numbers, and manufacturer. (Note: this does not include basic tools for checking grammar, spelling, references, etc.) |
| **Hepatology** | Medicine | Journal website | Yes | Authors who use AI tools in the writing of a manuscript, production of images or graphical elements of the paper, or in the collection and analysis of data, must be transparent in disclosing in the Materials and Methods (or similar section) of the paper how the AI tool was used and which tool was used. Authors are fully responsible for the content of their manuscript, even those parts produced by an AI tool, and are thus liable for any breach of publication ethics. | disclosing in the Materials and Methods (or similar section) of the paper how the AI tool was used and which tool was used. |
| **Lancet Oncology** | Medicine | both | Yes | Where authors use AI and AI-assisted technologies in the writing<br>process, these technologies should only be used to improve<br>readability and language of the work and not used to replace<br>researcher tasks such as producing scientific insights, analysing<br>and interpreting data, or drawing scientific conclusions. Applying<br>these technologies should only be done with human oversight and<br>control, and authors should carefully review and edit the result<br>because AI can generate authoritative-sounding output that can<br>be incorrect, incomplete, or biased. Authors should not list AI and<br>AI-assisted technologies as an author or co-author, nor cite AI as an<br>author. Authors are ultimately responsible and accountable for the<br>originality, accuracy, and integrity of the work; and should disclose<br>the use of AI and AI-assisted technologies in a statement at the end<br>of the article. | Authors... should disclose<br>the use of AI and AI-assisted technologies in a statement at the end<br>of the article. |
| **American Journal of Psychiatry** | Medicine | N/A | No | N/A | N/A |
| **Plant Cell** | Agriculture and Biological Sciences, Biochemistry, Genetics and Molecular Biology | Publisher website | No | N/A | N/A |
| **Journal of Chemical Physics** | Chemistry, Medicine, Physics and Astronomy | Publisher website | yes | ChatGPT and similar AI-based large language models should not be listed as an author. As with other instrumentation and software, the use of AI-based large language models such as ChatGPT should be disclosed to editors and reviewers, particularly if they are used to generate significant amounts of text in the manuscript. Authors should provide this information in the appropriate section of their manuscript and to the editor with their submission. | ChatGPT should be disclosed to editors and reviewers, particularly if they are used to generate significant amounts of text in the manuscript. Authors should provide this information in the appropriate section of their manuscript and to the editor with their submission. |
| **Cancer Cell** | Biochemistry, Genetics and Molecular Biology, Medicine | both | Yes | The below guidance only refers to the writing process, and not to the use of AI tools to analyze and draw insights from data as part of the research process.<br><br>Where authors use generative artificial intelligence (AI) and AI-assisted technologies in the writing process, authors should only use these technologies to improve readability and language. Applying the technology should be done with human oversight and control, and authors should carefully review and edit the result, as AI can generate authoritative-sounding output that can be incorrect, incomplete or biased. AI and AI-assisted technologies should not be listed as an author or co-author, or be cited as an author. Authorship implies responsibilities and tasks that can only be attributed to and performed by humans, as outlined in Elsevier's AI policy for authors.<br><br>Authors should disclose in their manuscript the use of AI and AI-assisted technologies in the writing process by following the instructions below. A statement will appear in the published work. Please note that authors are ultimately responsible and accountable for the contents of the work.<br><br>Disclosure instructions<br>Authors must disclose the use of generative AI and AI-assisted technologies in the writing process by adding a statement at the end of their manuscript. The statement should be placed in a new section after the 'Declaration of interests' section and the optional 'Inclusion and diversity' section, entitled 'Declaration of Generative AI and AI-assisted technologies in the writing process'.<br><br>Statement: During the preparation of this work the author(s) used [NAME TOOL / SERVICE] in order to [REASON]. After using this tool/service, the author(s) reviewed and edited the content as needed and take(s) full responsibility for the content of the | Authors must disclose the use of generative AI and AI-assisted technologies in the writing process by adding a statement at the end of their manuscript. The statement should be placed in a new section after the 'Declaration of interests' section and the optional 'Inclusion and diversity' section, entitled 'Declaration of Generative AI and AI-assisted technologies in the writing process'.<br><br>Statement: During the preparation of this work the author(s) used [NAME TOOL / SERVICE] in order to [REASON]. After using this tool/service, the author(s) reviewed and edited the content as needed and take(s) full responsibility for the content of the publication.<br><br>This declaration does not apply to the use of basic tools for checking grammar, spelling, references etc. If there is nothing to disclose, there is no need to add a statement. |
| | | | | publication.<br><br>This declaration does not apply to the use of basic tools for checking grammar, spelling, references etc. If there is nothing to disclose, there is no need to add a statement. | |
| **Journal of Clinical Endocrinology and Metabolism** | Biochemistry, Genetics and Molecular Biology, Medicine | N/A | No | N/A | N/A |
| **Pediatrics** | Medicine | Journal website | Yes | Artificial intelligence (AI) tools do not qualify for authorship. To qualify, authors must meet all four of the following criteria1:<br><br>Substantial contribution(s) to conception and design, acquisition of data, or analysis and interpretation of data; and<br>Drafting the article or revising it critically for important intellectual content; and<br>Final approval of the version to be published, and<br>Agreement to be accountable for all aspects of the work in ensuring that questions related to the accuracy or integrity of any part of the work are appropriately investigated and resolved.<br><br>AI tools cannot take responsibility for the accuracy or integrity of a manuscript and, therefore, do not qualify for authorship.2<br><br>While the use of AI tools is discouraged, if generative AI tools are used in any part of manuscript preparation, from writing to data analysis to image creation, the authors must report it in the Methods and Acknowledgments sections3 and note use of an AI tool in the cover letter. Identification of AI must include the name and manufacturer of the AI tool and how it was used in relation to the work being submitted.2 Authors are accountable for the integrity and accuracy of all material in their manuscript, including any content generated by AI. | While the use of AI tools is discouraged, if generative AI tools are used in any part of manuscript preparation, from writing to data analysis to image creation, the authors must report it in the Methods and Acknowledgments sections3 and note use of an AI tool in the cover letter. Identification of AI must include the name and manufacturer of the AI tool and how it was used in relation to the work being submitted. |
| **Physical Review D** | Physics and Astronomy | N/A | No | N/A | N/A |
| **Renewable and Sustainable Energy Reviews** | Energy | both | Yes | The below guidance only refers to the writing process, and not to the use of AI tools to analyze and draw insights from data as part of the research process.<br><br>Where authors use generative artificial intelligence (AI) and AI-assisted technologies in the writing process, authors should only use these technologies to improve readability and language. Applying the technology should be done with human oversight and control, and authors should carefully review and edit the result, as AI can generate authoritative-sounding output that can be incorrect, incomplete or biased. AI and AI-assisted technologies should not be listed as an author or co-author, or be cited as an author. Authorship implies responsibilities and tasks that can only be attributed to and performed by humans, as outlined in Elsevier's AI policy for authors.<br><br>Authors should disclose in their manuscript the use of AI and AI-assisted technologies in the writing process by following the instructions below. A statement will appear in the published work. Please note that authors are ultimately responsible and accountable for the contents of the work.<br><br>Disclosure instructions<br>Authors must disclose the use of generative AI and AI-assisted technologies in the writing process by adding a statement at the end of their manuscript. The statement should be placed in a new section after the 'Declaration of interests' section and the optional 'Inclusion and diversity' section, entitled 'Declaration of Generative AI and AI-assisted technologies in the writing process'.<br><br>Statement: During the preparation of this work the author(s) used [NAME TOOL / SERVICE] in order to [REASON]. After using this tool/service, the author(s) reviewed and edited the content as needed and take(s) full responsibility for the content of the | Authors must disclose the use of generative AI and AI-assisted technologies in the writing process by adding a statement at the end of their manuscript. The statement should be placed in a new section after the 'Declaration of interests' section and the optional 'Inclusion and diversity' section, entitled 'Declaration of Generative AI and AI-assisted technologies in the writing process'.<br><br>Statement: During the preparation of this work the author(s) used [NAME TOOL / SERVICE] in order to [REASON]. After using this tool/service, the author(s) reviewed and edited the content as needed and take(s) full responsibility for the content of the publication.<br><br>This declaration does not apply to the use of basic tools for checking grammar, spelling, references etc. If there is nothing to disclose, there is no need to add a statement. |
| | | | | publication.<br><br>This declaration does not apply to the use of basic tools for checking grammar, spelling, references etc. If there is nothing to disclose, there is no need to add a statement. | |
| **Journal of the National Cancer Institute** | Biochemistry, Genetics and Molecular Biology, Medicine | both | Yes | Natural language processing tools driven by artificial intelligence (AI) do not qualify as authors, and the Journal will screen for them in author lists. The use of AI (for example, to help generate content, write code, or process data) should be disclosed both in cover letters to editors and in the Methods or Acknowledgements section of manuscripts. Please see the COPE position statement on Authorship and AI for more details. | The use of AI (for example, to help generate content, write code, or process data) should be disclosed both in cover letters to editors and in the Methods or Acknowledgements section of manuscripts. |
| **Advanced Functional Materials** | Chemistry, Material Science, Physics and Astronomy | Publisher website | Yes | Artificial Intelligence Generated Content (AIGC) tools—such as ChatGPT and others based on large language models (LLMs)—cannot be considered capable of initiating an original piece of research without direction by human authors. They also cannot be accountable for a published work or for research design, which is a generally held requirement of authorship (as discussed in the previous section), nor do they have legal standing or the ability to hold or assign copyright. Therefore—in accordance with COPE's position statement on AI tools—these tools cannot fulfill the role of, nor be listed as, an author of an article. If an author has used this kind of tool to develop any portion of a manuscript, its use must be described, transparently and in detail, in the Methods or Acknowledgements section. The author is fully responsible for the accuracy of any information provided by the tool and for correctly referencing any supporting work on which that information depends. Tools that are used to improve spelling, grammar, and general editing are not included in the scope of these guidelines. The final decision about whether use of an AIGC tool is appropriate or permissible in the circumstances of a submitted manuscript or a published article lies with the journal's editor or other party responsible for the publication's editorial policy. | If an author has used this kind of tool to develop any portion of a manuscript, its use must be described, transparently and in detail, in the Methods or Acknowledgements section. |
| **JAMA Internal Medicine** | Medicine | Journal website | Yes | Reproduced and Recreated Material and Image Integrity: The submission and publication of content created by artificial intelligence, language models, machine learning, or similar technologies is discouraged, unless part of formal research design or methods, and is not permitted without clear description of the content that was created and the name of the model or tool, version and extension numbers, and manufacturer. Authors must take responsibility for the integrity of the content generated by these models and tools... Authorship Criteria and Contributions: Nonhuman artificial intelligence, language models, machine learning, or similar technologies do not qualify for authorship. If these models or tools are used to create content or assist with writing or manuscript preparation, authors must take responsibility for the integrity of the content generated by these tools. Authors should report the use of artificial intelligence, language models, machine learning, or similar technologies to create content or assist with writing or editing of manuscripts in the Acknowledgment section or Methods section if this is part of formal research design or methods... Acknowledgement Section: Authors should report the use of artificial intelligence, language models, machine learning, or similar technologies to create content or assist with writing or editing of manuscripts in the Acknowledgment section or the Methods section if this is part of formal research design or methods. This should include a description of the content that was created or edited and the name of the language model or tool, version and extension numbers, and manufacturer. (Note: this does not include basic tools for checking grammar, spelling, references, etc.) | Authors should report the use of artificial intelligence, language models, machine learning, or similar technologies to create content or assist with writing or editing of manuscripts in the Acknowledgment section or the Methods section if this is part of formal research design or methods. This should include a description of the content that was created or edited and the name of the language model or tool, version and extension numbers, and manufacturer. (Note: this does not include basic tools for checking grammar, spelling, references, etc.) |
| **Physiological Reviews** | Biochemistry, Genetics and Molecular Biology, Medicine | N/A | No | N/A | N/A |
| **Clinical Infectious Diseases** | Medicine | Journal website | Yes | CID supports the World Association of Medical Editors’ recommendations on chatbots and scholarly manuscripts. If a chatbot or similar program is used in the development of a paper for CID, the following is required:<br><br>The Large Language Models (LLM) cannot be credited as an author, as authorship requires that the author be accountable for the submitted/published work, and artificial intelligence cannot fulfill this requirement of authorship.<br>Authors listed on the paper must review the content generated by the LLM and take full responsibility for it, as they would for any other content within the submitted/published work.<br>The use of LLM tools must be noted in the cover letter.<br>The use of LLM tools must be documented in the Methods, Acknowledgments, or another appropriate section of the paper. | The use of LLM tools must be noted in the cover letter.<br>The use of LLM tools must be documented in the Methods, Acknowledgments, or another appropriate section of the paper. |
| **Reviews of Modern Physics** | Physics and Astronomy | N/A | No | N/A | N/A |
| **Nature Reviews Drug Discovery** | Medicine, Pharmacology, Toxicology and Pharmaceutics | Journal website | Yes | Large Language Models (LLMs), such as ChatGPT, do not currently satisfy our authorship criteria. Notably an attribution of authorship carries with it accountability for the work, which cannot be effectively applied to LLMs. Use of an LLM should be properly documented in the Methods section (and if a Methods section is not available, in a suitable alternative part) of the manuscript. | Use of an LLM should be properly documented in the Methods section (and if a Methods section is not available, in a suitable alternative part) of the manuscript. |
| **Trends in Ecology and Evolution** | Agriculture and Biological Sciences | both | Yes | The below guidance only refers to the writing process, and not to the use of AI tools to analyze and draw insights from data as part of the research process.<br><br>Where authors use generative artificial intelligence (AI) and AI-assisted technologies in the writing process, authors should only use these technologies to improve readability and language. Applying the technology should be done with human oversight and control, and authors should carefully review and edit the result, as AI can generate authoritative-sounding output that can be incorrect, incomplete or biased. AI and AI-assisted technologies should not be listed as an author or co-author, or be cited as an author. Authorship implies responsibilities and tasks that can only be attributed to and performed by humans, as outlined in Elsevier's AI policy for authors.<br><br>Authors should disclose in their manuscript the use of AI and AI-assisted technologies in the writing process by following the instructions below. A statement will appear in the published work. Please note that authors are ultimately responsible and accountable for the contents of the work.<br><br>Disclosure instructions<br>Authors must disclose the use of generative AI and AI-assisted technologies in the writing process by adding a statement at the end of their manuscript. The statement should be placed in a new section after the 'Declaration of interests' section and the optional 'Inclusion and diversity' section, entitled 'Declaration of Generative AI and AI-assisted technologies in the writing process'.<br><br>Statement: During the preparation of this work the author(s) used [NAME TOOL / SERVICE] in order to [REASON]. After using this tool/service, the author(s) reviewed and edited the content as needed and take(s) full responsibility for the content of the | Authors must disclose the use of generative AI and AI-assisted technologies in the writing process by adding a statement at the end of their manuscript. The statement should be placed in a new section after the 'Declaration of interests' section and the optional 'Inclusion and diversity' section, entitled 'Declaration of Generative AI and AI-assisted technologies in the writing process'.<br><br>Statement: During the preparation of this work the author(s) used [NAME TOOL / SERVICE] in order to [REASON]. After using this tool/service, the author(s) reviewed and edited the content as needed and take(s) full responsibility for the content of the publication.<br><br>This declaration does not apply to the use of basic tools for checking grammar, spelling, references etc. If there is nothing to disclose, there is no need to add a statement. |
| | | | | publication.<br><br>This declaration does not apply to the use of basic tools for checking grammar, spelling, references etc. If there is nothing to disclose, there is no need to add a statement. | |
| **Circulation Research** | Biochemistry, Genetics and Molecular Biology, Medicine | Journal website | Yes | The use of automated assistive writing technologies and tools (commonly referred to as artificial intelligence or machine learning tools) is permitted provided that their use is documented and authors assume responsibility for the content. As with human-generated content, authors are responsible for the accuracy, validity, and originality of computer-generated content. Per ICMJE Authorship Criteria, automated assistive writing technologies do not qualify for authorship as they are unable to provide approval or consent for submission. Per ICMJE recommendations for writing assistance, these tools should be listed in the Acknowledgements; if involved in the research design, the tools should be documented in the Methods. For additional information, see the World Association of Medical Editor recommendations. | Per ICMJE recommendations for writing assistance, these tools should be listed in the Acknowledgements; if involved in the research design, the tools should be documented in the Methods. |
| **American Journal of Clinical Nutrition** | Medicine, Nursing | N/A | No | N/A | N/A |
| **Brain** | Medicine | Publisher website | No | N/A | N/A |
| **Chemical Communications** | Chemical Engineering, Chemistry, Materials Science | Publisher website | Yes | Artificial intelligence (AI) tools, such as ChatGPT or other Large Language Models, cannot be listed as an author on a submitted work. AI tools do not meet the criteria to qualify for authorship, as they are unable to take responsibility for the work, cannot consent to publication nor manage copyright, licence or other legal obligations, and are unable to understand issues around conflicts of interest. Any use of AI tools in producing any part of the manuscript must be clearly described in the Experimental or Acknowledgement section. The authors are fully responsible and accountable for the content of their article, including any parts produced by an AI tool. | Any use of AI tools in producing any part of the manuscript must be clearly described in the Experimental or Acknowledgement section. |
| **Nature Methods** | Biochemistry, Genetics and Molecular Biology | Journal website | Yes | Large Language Models (LLMs), such as ChatGPT, do not currently satisfy our authorship criteria. Notably an attribution of authorship carries with it accountability for the work, which cannot be effectively applied to LLMs. Use of an LLM should be properly documented in the Methods section (and if a Methods section is not available, in a suitable alternative part) of the manuscript. | Use of an LLM should be properly documented in the Methods section (and if a Methods section is not available, in a suitable alternative part) of the manuscript. |
| **Clinical Cancer Research** | Biochemistry, Genetics and Molecular Biology, Medicine | N/A | No | N/A | N/A |
| **Nature Photonics** | Materials Science, Physics and Astronomy | Journal website | Yes | Large Language Models (LLMs), such as ChatGPT, do not currently satisfy our authorship criteria. Notably an attribution of authorship carries with it accountability for the work, which cannot be effectively applied to LLMs. Use of an LLM should be properly documented in the Methods section (and if a Methods section is not available, in a suitable alternative part) of the manuscript. | Use of an LLM should be properly documented in the Methods section (and if a Methods section is not available, in a suitable alternative part) of the manuscript. |
| **Oncogene** | Biochemistry, Genetics and Molecular Biology | Jounal | Yes | Large Language Models (LLMs), such as ChatGPT, do not currently satisfy our authorship criteria. Notably an attribution of authorship carries with it accountability for the work, which cannot be effectively applied to LLMs. Use of an LLM should be properly documented in the Methods section (and if a Methods section is not available, in a suitable alternative part) of the manuscript. | Use of an LLM should be properly documented in the Methods section (and if a Methods section is not available, in a suitable alternative part) of the manuscript. |
| **Diabetes** | Medicine | Journal website | Yes | ADA has adopted and modified JAMA's instructions for authors to address the roles of artificial intelligence (AI) and machine learning in the development of content to be presented in ADA publications. See below for more information related to authorship, image integrity, and reproduced and re-created material.<br><br>AI, Authorship, and Content Creation. Nonhuman artificial intelligence, language models, machine learning, or similar technologies do not qualify for authorship.<br><br>If these models or tools are used to create content or assist with writing or manuscript preparation, authors must take responsibility for the integrity of the content generated by these tools.<br><br>Authors should report the use of artificial intelligence, language models, machine learning, or similar technologies to create content or assist with writing or editing of manuscripts in the Acknowledgments section or the Methods section if this is part of formal research design or methods, as well as in the comments to the editors at the time of submission.<br><br>This should include a description of the content that was created or edited and the name of the language model or tool, version and extension numbers, manufacturer, and (where relevant) the query or prompt to create or assist with the development of content. (Note: this does not include basic tools for checking grammar, spelling, references, etc.).<br><br>AI and Image Development. The submission and publication of images created by artificial intelligence, machine learning tools, or similar technologies is discouraged, unless part of formal research design or methods, and is not permitted without clear description of the content that was created and the name of the model or tool, version | Authors should report the use of artificial intelligence, language models, machine learning, or similar technologies to create content or assist with writing or editing of manuscripts in the Acknowledgments section or the Methods section if this is part of formal research design or methods, as well as in the comments to the editors at the time of submission.<br><br>This should include a description of the content that was created or edited and the name of the language model or tool, version and extension numbers, manufacturer, and (where relevant) the query or prompt to create or assist with the development of content. (Note: this does not include basic tools for checking grammar, spelling, references, etc.). |
| | | | | and extension numbers, manufacturer, and (where relevant) the query or prompt to create or assist with the development of content. Authors must take responsibility for the integrity of the content generated by these models and tools.<br><br>AI and Reproduced Material. The submission and publication of content created by artificial intelligence, language models, machine learning, or similar technologies is discouraged, unless part of formal research design or methods, and is not permitted without clear description of the content that was created and the name of the model or tool, version and extension numbers, manufacturer, and (where relevant) the query or prompt to create or assist with the development of content. Authors must take responsibility for the integrity of the content generated by these models and tools. | |
| **Langmuir** | Chemistry, Material Science, Medicine, Physics and Astronomy | Publisher website | Yes | Artificial intelligence (AI) tools do not qualify for authorship. The use of AI tools for text or image generation should be disclosed in the manuscript within the Acknowledgment section with a description of when and how the tools were used. For more substantial use cases or descriptions of AI tool use, authors should provide full details within the Methods or other appropriate section of the manuscript. | The use of AI tools for text or image generation should be disclosed in the manuscript within the Acknowledgment section with a description of when and how the tools were used. For more substantial use cases or descriptions of AI tool use, authors should provide full details within the Methods or other appropriate section of the manuscript. |
| **Academy of Management Journal** | Business, Management, and Accounting | N/A | No | N/A | N/A |
| **Monthly Notices of the Royal Astronomical Society** | Earth and Planetary Sciences, Physics and Astronomy | Publisher website | No | N/A | N/A |
| **Analytical Chemistry** | Chemistry | Publisher website | Yes | Artificial intelligence (AI) tools do not qualify for authorship. The use of AI tools for text or image generation should be disclosed in the manuscript within the Acknowledgment section with a description of when and how the tools were used. For more substantial use cases or descriptions of AI tool use, authors should provide full details within the Methods or other appropriate section of the manuscript. | The use of AI tools for text or image generation should be disclosed in the manuscript within the Acknowledgment section with a description of when and how the tools were used. For more substantial use cases or descriptions of AI tool use, authors should provide full details within the Methods or other appropriate section of the manuscript. |